\documentclass[twocolumn]{aastex701}

\begin{document}

\title{Properties of Circumfacular Regions derived from H$\alpha$ and \ion{Ca}{2}~K  synoptic observations of the Sun}

\author[0000-0002-4525-9038]{Serena Criscuoli}
\affiliation{National Solar Observatory (NSO), 
             3665 Discovery Drive,
             Boulder, CO, USA, 80303}
\email[show]{scriscuo@nso.edu}

\author[0000-0002-9858-0490]{Andrea Diercke}
\affiliation{Institut f\"ur Sonnenphysik (KIS), 
            Georges-K\"ohler-Allee 401 A,
            79110 Freiburg, Germany}
\email{andrea.diercke@googlemail.com}


\begin{abstract}
Circumfacular regions are dark chromospheric structures surrounding magnetic regions whose properties and contribution to solar activity diagnostics remain poorly understood. We investigate their photometric and geometric properties using full-disk H$\alpha$ and \ion{Ca}{2}~K observations acquired by the Chromospheric Telescope (ChroTel) from 2012 to 2020, spanning the maximum and declining phases of solar cycle 24 and the beginning of cycle 25. We develop an automated segmentation algorithm to identify circumfacular regions and define a photometric index analogous to existing excess and deficit indices. We compare their temporal evolution with those of plages, dark features, and filaments on solar-cycle and Carrington-rotation timescales. Circumfacular regions develop after the H$\alpha$ plage brightening become visible and persist into the decay phase of active regions. At activity maximum, they are the most extended chromospheric structures, covering more than half of the visible solar disk. Their mean intensity varies only weakly throughout the solar cycle; consequently, their index variability, as well as that of the other photometric indices, is driven primarily by changes in area coverage. H$\alpha$ and \ion{Ca}{2}~K indices are strongly correlated, although their correlations weaken toward solar minimum. This activity-dependent relationship is primarily driven by the decline of the area of H$\alpha$ excess regions during periods of low activity, and the increasing relative contributions of deficit and circumfacular regions.  Our results therefore suggest that circumfacular regions should be considered when interpreting disk-integrated chromospheric observations of the Sun and solar-type stars.

\end{abstract}



\section{Introduction} \label{sec:intro}
Chromospheric and transition region observations of the Sun reveal that active regions, which typically appear bright in the higher layers of the solar atmosphere, are often surrounded by dark halos. These regions were first discovered by \citet{hale1903} in \ion{Ca}{2} K observations and were named "circumfaculairs" by D'Azambuja and Deslandres, as noted by \citet{bumba1965}. Later, they have been also addressed as "dark canopy" \citep{wang2011}, "dark halos" \citep{andretta2014} and "dark moats" \citep{singh2021} when observed in EUV wavelengths. \citet{bumba1965} were the first to note a connection between dark fibrils observed in H$\alpha$ observations and circumfaculae observed in \ion{Ca}{2} K observations acquired at the Mount Wilson Observatory. By analyzing the evolution of 152 active regions, they noted that a "dark ellipse" is visible in both \ion{Ca}{2} and H$\alpha$ observations, but that the ones observed in Calcium are somewhat smaller, and more visible at the peak of the \ion{Ca}{2} emission of the ARs plage evolution.  Subsequent observations showed that in circumfacular regions the core of \ion{Ca}{2} 854.2 nm line is depressed and slightly blueshifted with respect to quiet Sun \citep[e.g.][]{cauzzi2008, pietarila2013}. In particular, \citet{pietarila2013} employed full-disk spectropolarimetric observations in \ion{Ca}{2} 854.2 nm  acquired with VSM at SOLIS and GONG observations in H$\alpha$ to show that in circumfacular regions the \ion{Ca}{2} line bisector and the 3 minutes power in H$\alpha$ are suppressed. They concluded that this result is evidence of magnetic canopy suppressing acoustic waves propagating upward in the solar atmosphere. Complementary work by \citet{wang2011}, using AIA 171 $\AA$ and HMI magnetograms,
showed that dark fibrils preferentially form in zones of weak or mixed-polarity longitudinal magnetic
field. This was attributed to the cancellation of small-scale magnetic loops, leading fibrils to concentrate
near small-scale polarity inversion lines. Later, \citet{lezzi2023} showed that circumfacular regions are also clearly visible in IRIS data as fibrillar patterns seen in Mg II h $\&$ k line cores, and that the
associated EUV dark halo has generally a much larger area.

Despite these advances, the origin of circumfacular regions and Dark Halos remains unclear. As pointed out in \citet{andretta2014} and later in \citet{lezzi2023}, none of the models presented so far in the literature can explain why dark halos are observed in both chromospheric and transition region layers. The fact that they are observed at wavelengths shorter than Ly$\alpha$ indicates that these dark features are not manifestations of absorption, mainly from Hydrogen and Helium, which was the working assumption of \citet{wang2011}.    Later, \citet{singh2021} proposed that "dark moats" are caused by strong magnetic fields from active regions compressing magnetic loops to low altitudes, preventing the emission of 171 $\AA$ plasma. However, \citet{lezzi2023} noted that this model is inconsistent with observations of dark halos observed in 304 $\AA$ images. In order to better understand the physical mechanisms that produce the formation of circumfacular regions in different layers of the solar atmosphere, \citet{lezzi2024} analyzed data acquired at high spatial resolution with instrumentation aboard of the Solar Orbiter in various EUV lines, including Ly$\alpha$ at 121 nm and \ion{Fe}{9} and \ion{Fe}{10} in the 171 $\AA$ range, and photospheric magnetograms. They found that circumfacular regions observed in EUV imagery appear as fine structured, with bright regions (bundles) alternating to elongated dark filamentary features, that seem to originate from photospheric magnetic features.

In this work we compare photometric and geometric properties of circumfacular regions observed in H$\alpha$ and \ion{Ca}{2}~K using data acquired with the ChroTel instrument in solar cycle 24. We also define a circumfacular regions index, and perform a comparison between its evolution and the evolution of other indices also extracted from ChroTel observations. The methodology and the analysis are similar to the ones presented in \citet{diercke2022}, who showed that H$\alpha$ excess index (computed from bright regions - plages) and the deficit index (computed from dark features) extracted from ChroTel observations are good tracers of solar activity. Our analysis contributes to understanding the complex relation observed between the variability of indices extracted from disk-integrated observations of H$\alpha$ and \ion{Ca}{2}~K. Indeed, as fist noted by \citet{meunier2009} who analyzed Kitt Peak Observations, and later confirmed by \citet{criscuoli2023} from the analysis of ISS at SOLIS data, the correlation between H$\alpha$ and \ion{Ca}{2}~K indices changes with the level of activity, being larger during the ascending/maximum phase, and smaller during periods of minima. While a variation in phase of H$\alpha$ index with the activity was found by \citet{livingston2007,meunier2009,criscuoli2023, zills2024}, an anti-correlation was reported by \citet{maldonado2019} and \citet{toriumi2022}. A complex relation between H$\alpha$ and \ion{Ca}{2}~K emission was also found for solar-like stars \citep[e.g.][]{cincunegui2007,meunier2022,loaiza2025}, as well as for other lines of the Hydrogen Balmer series \citep[e.g.][]{flores2018, marchenko2021,dravins2024, he2024}. The presence of filaments, which decreases the emission of the H$\alpha$ core, is often invoked to explain the decrease of correlation, or the anti-correlation between H$\alpha$ and \ion{Ca}{2}~K indices. However, the semi-empirical model presented in \citet{criscuoli2023} indicates that filaments are unlikely to produce an anti-correlation between H$\alpha$ and other activity indices. On the other hand, as will be shown in the paper, circumfacular regions are extended regions, whose area occupies much larger fraction of the disk than the one occupied by filaments, and that surround active regions for most of their lives, so that they are more likely to cause a decrease of correlation of H$\alpha$ with other activity indices.   

The paper is organized as follows. Sections 2 to 4 describe the observations, the feature-identification algorithm, and the definition of the photometric indices. The observational results are presented in Section 5 and discussed in the context of chromospheric activity diagnostics and the H$\alpha$–\ion{Ca}{2}~K relationship in Section 6. The main conclusions are summarized in Section 7.

\section{Observations and data processing} \label{sec:data}

In this study, we focus the analysis on data from the Chromospheric Telescope \citep[ChroTel, ][]{kentischer2008, halbgewachs2008, bethge2011} located at the Observatorio del Teide in Tenerife, Spain. This ground-based robotic telescope, with a 10-centimeter aperture,  was operated by the Institute for Solar Physics (KIS) in Freiburg, Germany. It acquired full-disk observations of the Sun in three different chromospheric wavelengths  \citep{bethge2011}: H$\alpha$ at $\lambda$~6562.8\,\AA\ (Full Width at Half Maximum, FWHM = 0.5\,\AA), \ion{Ca}{2}~K at $\lambda$~3933.7\,\AA\ (FWHM = 0.3\,\AA), and \ion{He}{1} around $\lambda$~10830.3\,\AA\ (FWHM = 1.3\,\AA). In the standard operation mode, ChroTel observed each wavelength every three minutes, whereby the recorded data had a size of $2048 \times 2048$ pixels. The basic data processing including flat-field and dark correction was performed automatically on-site, so that the high-resolution observatories VTT and GREGOR could use its data as context data during observations. This Level 1.0 data is available for 2012 until 2020 in the Science Data Centre \citep[SDC, ][]{caligari2024} of KIS\footnote{KIS Science Data Centre: \href{https://archive.sdc.leibniz-kis.de/}{archive.sdc.leibniz-kis.de}}. The temporal coverage of the ChroTel observations is not uniform, owing primarily to weather conditions and instrumental downtime. The number of observing days per year is reported in Table 1 of \citet{diercke2022}.
We selected the qualitatively best image per day, resulting in a total of 1057 images in H$\alpha$ and 1010 in \ion{Ca}{2}~K.  We then applied  additional processing steps, including correction for differential refraction effects, limb-darkening correction, and correction of non-uniform intensity patterns with an approximation using Zernike polynomials \citep{shen2018}.  All images were scaled so that the solar radius is equal to $r_\odot=1000$\,pixels, which results in an image scale of about 0\farcs96\,pixel$^{-1}$. A more detailed description of the processing steps can be found in \citet{shen2018} and  \citet{diercke2019}. The non-uniform intensity corrected data is provided as ChroTel high-level data product in the SDC archive of KIS, as well. An overview of the acquired data between 2012 and 2018 is provided in Figure~1 in \citet{diercke2019}. The ChroTel observations started during the maximum of solar cycle~24 and cover the decline of solar activity, as well as the solar minimum in 2019/2020. As a result, the dataset provides a representative statistical sample of different activity phases of solar cycle~24. 
Additionally, we used filament masks publicly available at the KIS-SDC derived automatically from the ChroTel data using a semi-supervised learning neural network \citep{diercke2024}.

\begin{figure*}
\centering
\includegraphics[width=0.92\textwidth,trim={0 4cm 0 3.5cm},clip]{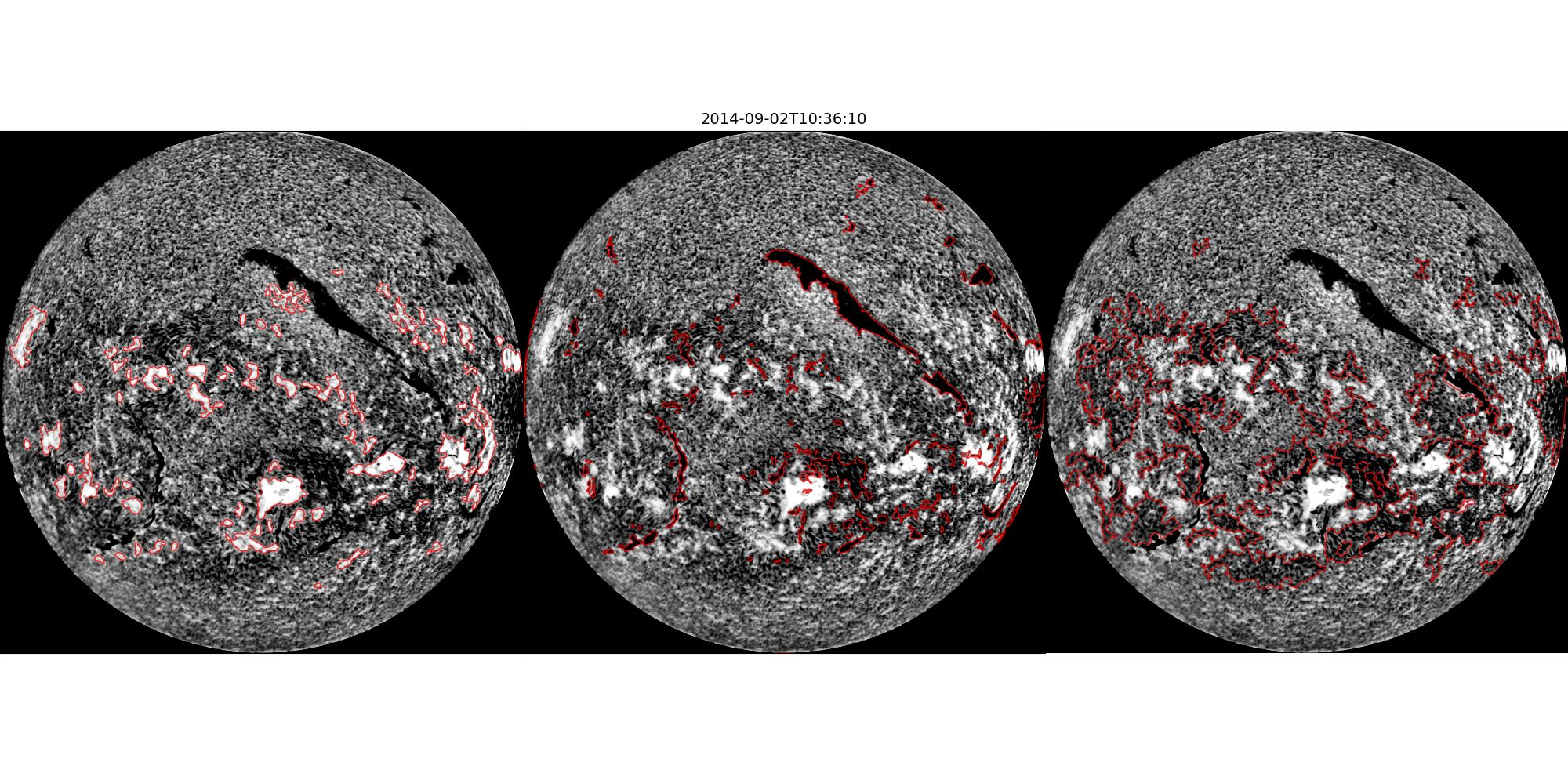}
\includegraphics[width=0.92\textwidth, trim={0 4.2cm 0 3cm},clip]
{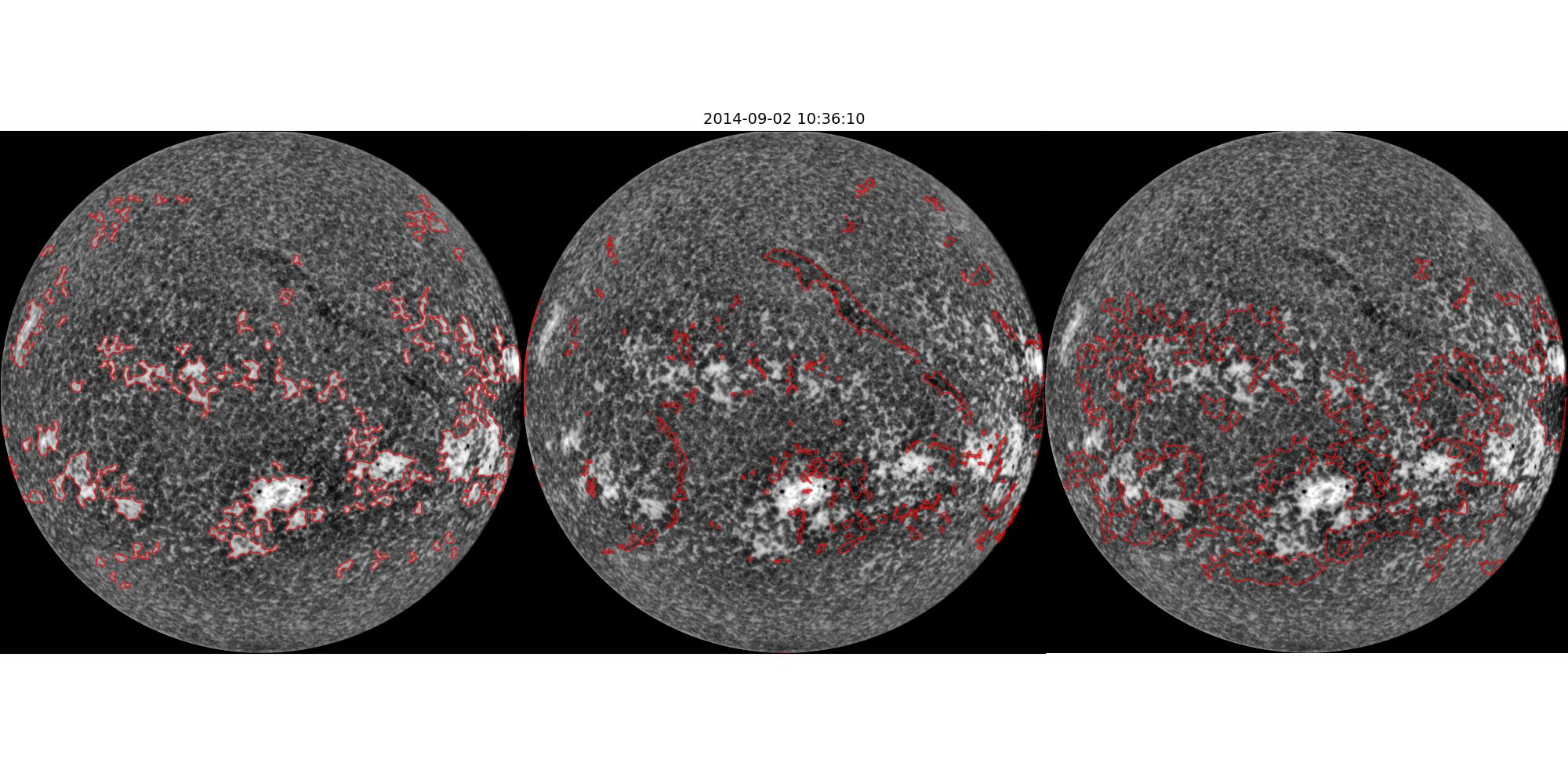}
\caption{Example images for H$\alpha$ (top) and \ion{Ca}{2}~K  (bottom) showing the segmentation of excess, deficit and circumfacular regions structures (red contours, from left to right).} 
\label{fig:index}
\end{figure*}

\begin{figure*}
\centering
\includegraphics[width=0.7\textwidth]{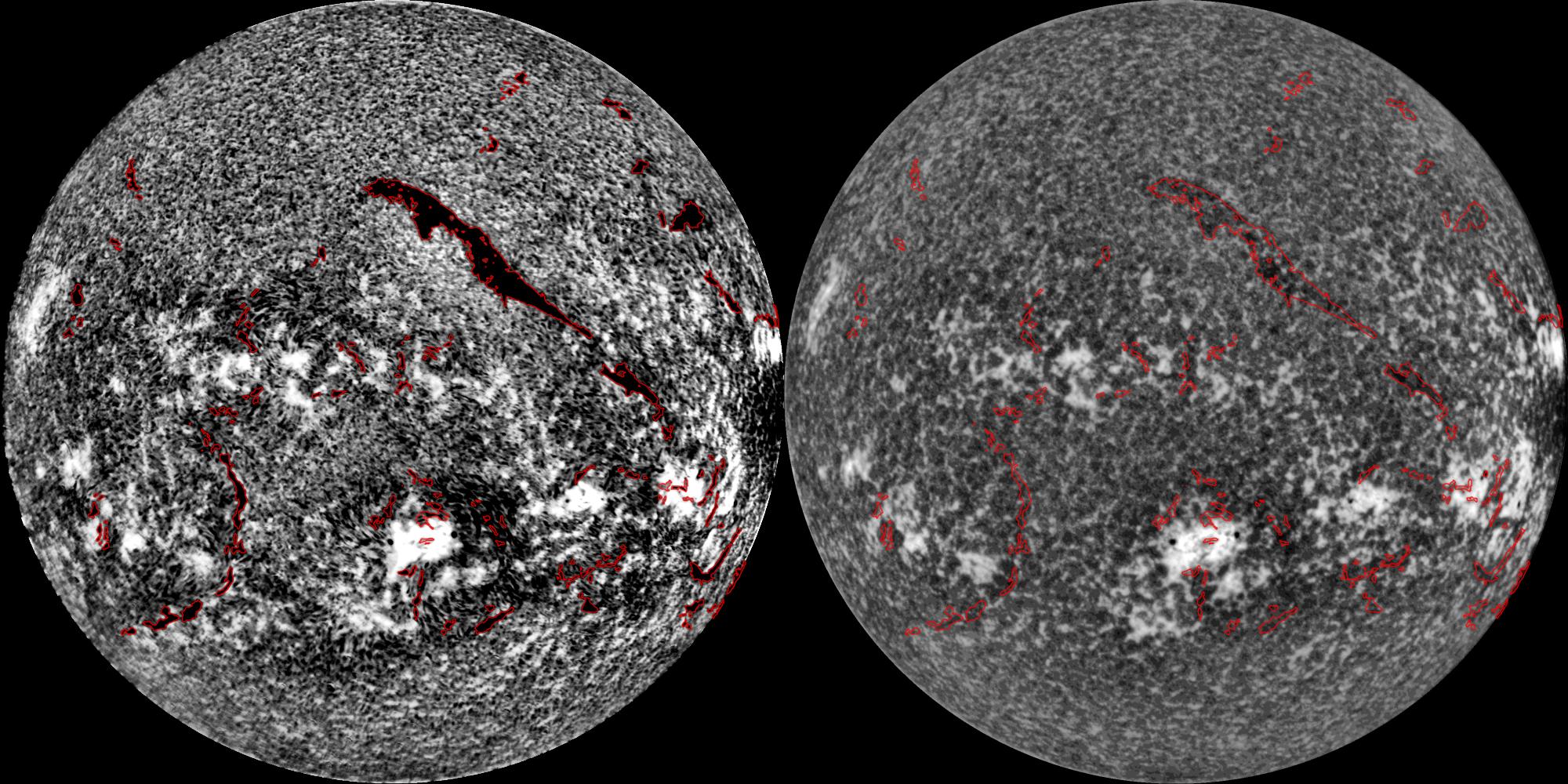}
\caption{Example image of filament segmentation. The contours of filaments (red lines) are overlaid to the H$\alpha$ (left) and \ion{Ca}{2}~K (right) images shown in Fig.~\ref{fig:index}.} 
\label{fig:overview_fil}
\end{figure*}

\section{Detection Algorithm} \label{sec:detection}

For the detection algorithm we use the intensity corrected Level~2.0 ChroTel data which are available in the SDC data archive. The data are already normalized to the median value of the quiet-Sun intensity $I_\mathrm{med}$. Additionally, we use adaptive histogram equalization to enhance the contrast of the filtergrams and improve the visibility of circumfacular regions (see Figure~\ref{fig:index}). 

The main focus of this study is on circumfacular regions, but we also compare these regions with excess, deficit and filament regions. Excess regions are bright plage regions in H$\alpha$ or \ion{Ca}{2}~K. Deficit regions are dark structures on the solar disk such as filaments, sunspots or larger regions of fibrils. In addition we distinguish from the deficit regions explicitly filaments as an additional subset. To extract the excess and deficit regions, we use a similar method as described in \citet{diercke2022}, but with different median values. The median values had to be adjusted because of the additional adaptive histogram equalization we use in this paper.

To extract the excess regions from the enhanced images, we select all pixels whose intensity is larger than $I_\mathrm{med} + 0.3 \cdot I_\mathrm{med}$, where $I_\mathrm{med}$ is the median value of each image. On these contours we use a morphological opening algorithm utilizing a disk kernel $k$ with a radius of six pixels. Afterwards the holes in the contour are filled and objects smaller than 500 pixels are removed. An additional exclusion of contours in the polar regions is applied in the end. Excess regions, mostly composed of plages, where selected separately on H$\alpha$ and \ion{Ca}{2}~K images. We note that this approach resulted in larger features than the ones found by \citet{diercke2022}.

The deficit regions are extracted from H$\alpha$ images with an analogous procedure, but utilizing a threshold at $I_\mathrm{med} - 0.4 \cdot I_\mathrm{med}$. The disk kernel has a kernel size of eight pixels. Furthermore, objects smaller than 50 pixels are removed from the contour. In addition, we use H$\alpha$ filament segmentation maps available at the KIS-SDC to include all filaments in the deficit extraction. These masks were derived from ChroTel H$\alpha$ observations using a semi-supervised learning approach based on a fully convolutional neural network, as described in \citet{diercke2024}. 

Filaments were also analyzed separately from the deficit regions, to better understand their properties and contributions to variability.

The circumfacular regions are morphologically darker regions around active regions which do not reach the absorption level of deficit regions. Therefore, we use a threshold with $I_\mathrm{med} - 0.15 \cdot I_\mathrm{med}$ on H$\alpha$ data. The following steps are similar to the extraction of the excess and deficit regions. We apply a morphological opening algorithm on the H$\alpha$ images with a disk kernel with a kernel size of 12 pixels. After filling holes, we remove objects smaller than 200 pixels, since we expect circumfacular regions to be large connected regions. In addition, the contours of excess and deficit regions, including the filament regions are excluded from the circumfacular region mask. Furthermore, we assume that circumfacular regions are located in the activity belt, whereby we exclude all objects close to the polar regions. 

For all features, the adopted thresholds were determined empirically through inspection of a representative sample of images, with the aim of obtaining a satisfactory identification of the different structures over a range of activity levels. As is generally the case for intensity-threshold segmentation, these values are not unique, and variations in the adopted thresholds affect the inferred feature areas and, consequently, the amplitudes of the corresponding photometric indices. This aspect will be further discussed in Sec.~\ref{sec:comp_diercke}.

In summary, while the masks for the excess regions were derived separately on H$\alpha$ and \ion{Ca}{2}~K images, deficit, cirumfacular and filaments were identified using H$\alpha$ images only, as these regions appear fainter in \ion{Ca}{2}~K data, and are harder to identify through an automatic detection algorithm. An example of the contours of excess, deficit and circumfacular regions, overlaid on H$\alpha$ and \ion{Ca}{2}~K images, is provided in Fig.~\ref{fig:index}. Figure~\ref{fig:overview_fil} shows, for the same date, the contours of filaments only.

\section{Photometric Indices}\label{sec: indices}

We use the masks produced by the segmentation algorithm described above to compute photometric indices, following the definitions adopted by \citet{diercke2022}, which, for the excess regions, followed the definitions of \citet{Johannesson1998} and \citet{Naqvi2010}. Briefly, for the excess regions, which are identified independently in H$\alpha$ and Ca II K, the index is computed as the sum of the absolute differences between the pixel intensities and the corresponding excess threshold ($I_{med}+0.3I_{med}$). Indices of other features are defined in a similar manner. Specifically, for the deficit regions the same H$\alpha$-derived mask is applied to both passbands. The H$\alpha$ deficit index is computed relative to the H$\alpha$ threshold, $T^{H\alpha}_{def.} = I^{H\alpha}_{med} - 0.4I^{H\alpha}_{med}$, whereas for \ion{Ca}{2}~K we define an analogous reference threshold from the \ion{Ca}{2}~K image, $T^{CaIIK}_{def.} = I^{CaIIK}_{med} - 0.4I^{CaIIK}_{med}$, and sum the absolute differences between the Ca II K intensities within the H$\alpha$-defined deficit mask $T^{CaIIK}$. The filament deficit indices are computed using the same thresholds ($I_{med}-0.4I_{med}$, computed separately for each passband), and using the filament masks derived from H$\alpha$ observations. For circumfacular regions, the indices are computed by using as intensity thresholds $T^{H\alpha}_{circ.} = I^{H\alpha}_{med} - 0.15I^{H\alpha}_{med}$ and $T^{CaIIK}_{circ.} = I^{CaIIK}_{med} - 0.15I^{CaIIK}_{med}$, on the two passbands, summing over pixels belonging to masks identified in H$\alpha$ observations. All indices are normalized by a factor of 1000.



\section{Results} \label{sec:results}

To better contextualize and compare our findings with existing literature, we first perform visual inspection of the obtained masks, and relate the evolution of the circumfacular regions with the evolution of the active regions. These results are presented in Sec.~\ref{sec:ratios}.
We then study the variation of the circumfacular and of the other indices at the solar cycle temporal scale, and on the Carrington rotation scale. The former was investigated analyzing the variations over the whole ChroTel database, therefore from the maximum of cycle 24 to the subsequent minimum and start of cycle 25, for a total of eight years. The Carrington scale was investigated by grouping the results over several Carrington rotations. Results from these analyses are presented in Sec.~\ref{sec:cycle} and Sec.~\ref{sec:activelong}, respectively.

\subsection{Evolution of Circumfacular Regions} \label{sec:ratios}

Visual inspection of a few relatively isolated active regions suggests that circumfacular regions generally become apparent after the associated H$\alpha$ plage brightening. In the cases that could be followed, this occurred approximately one to two days after the plage became visible, consistent with the results reported by \citet{bumba1965}. However, this estimate should be regarded as qualitative. A statistical determination of the delay is prevented by the limited temporal sampling and gaps in the ground-based observations, as well as by the difficulty of associating individual circumfacular regions with specific active regions, particularly during periods of high activity when neighboring circumfacular regions frequently merge.

\begin{figure*}
\centering
\includegraphics[width=0.9\textwidth,trim={3cm 43cm 2.cm 40cm},clip]{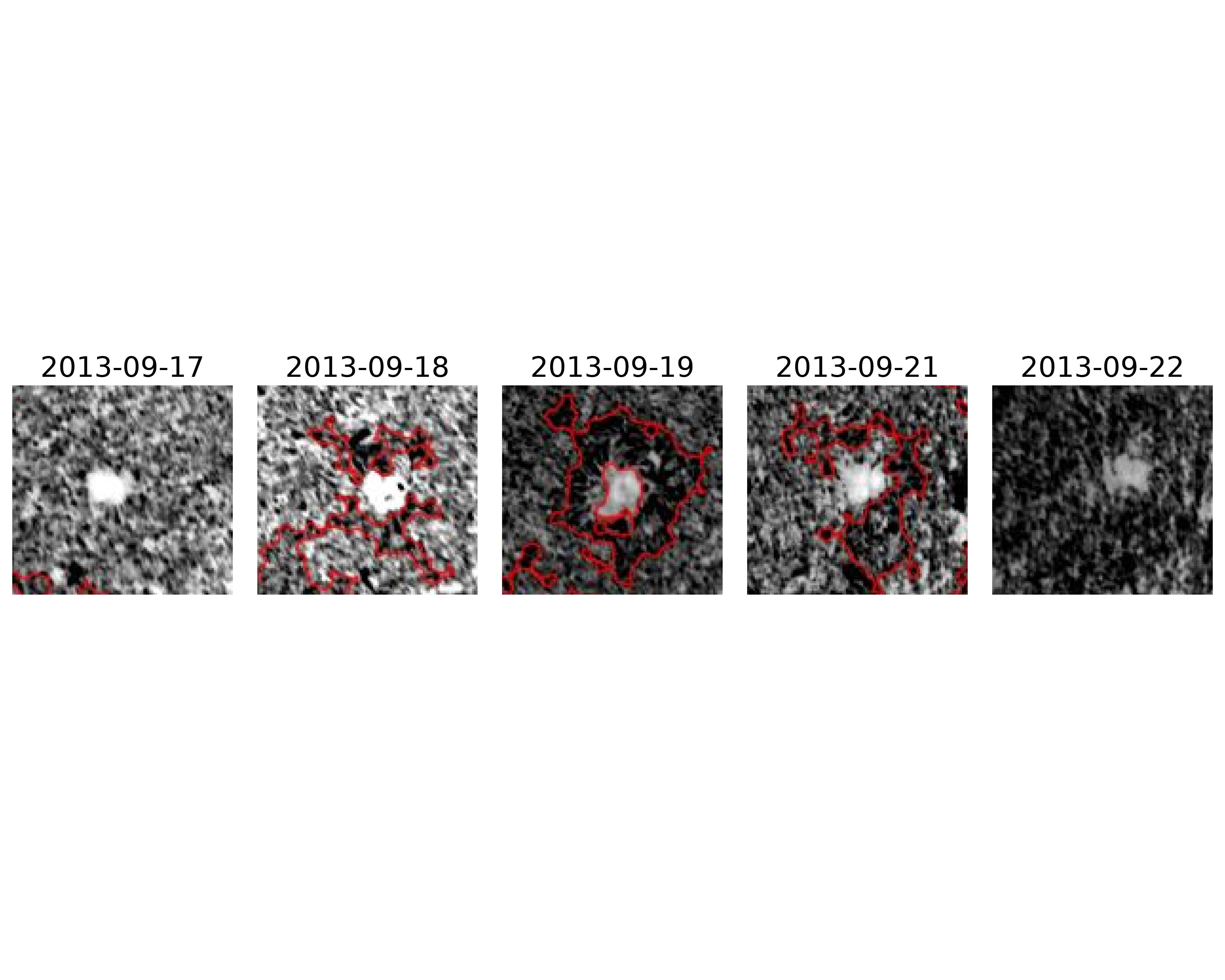}
\caption{Details of ChroTel H$\alpha$ images showing the evolution of AR11843 and its surrounding circumfacular region. The boundary of the circumfacular region is delineated by the red contour.} 
\label{fig:AR11843}
\end{figure*}

\begin{figure*}
\centering
\includegraphics[width=0.9\textwidth,trim={3cm 35cm 2.cm 30cm},clip]{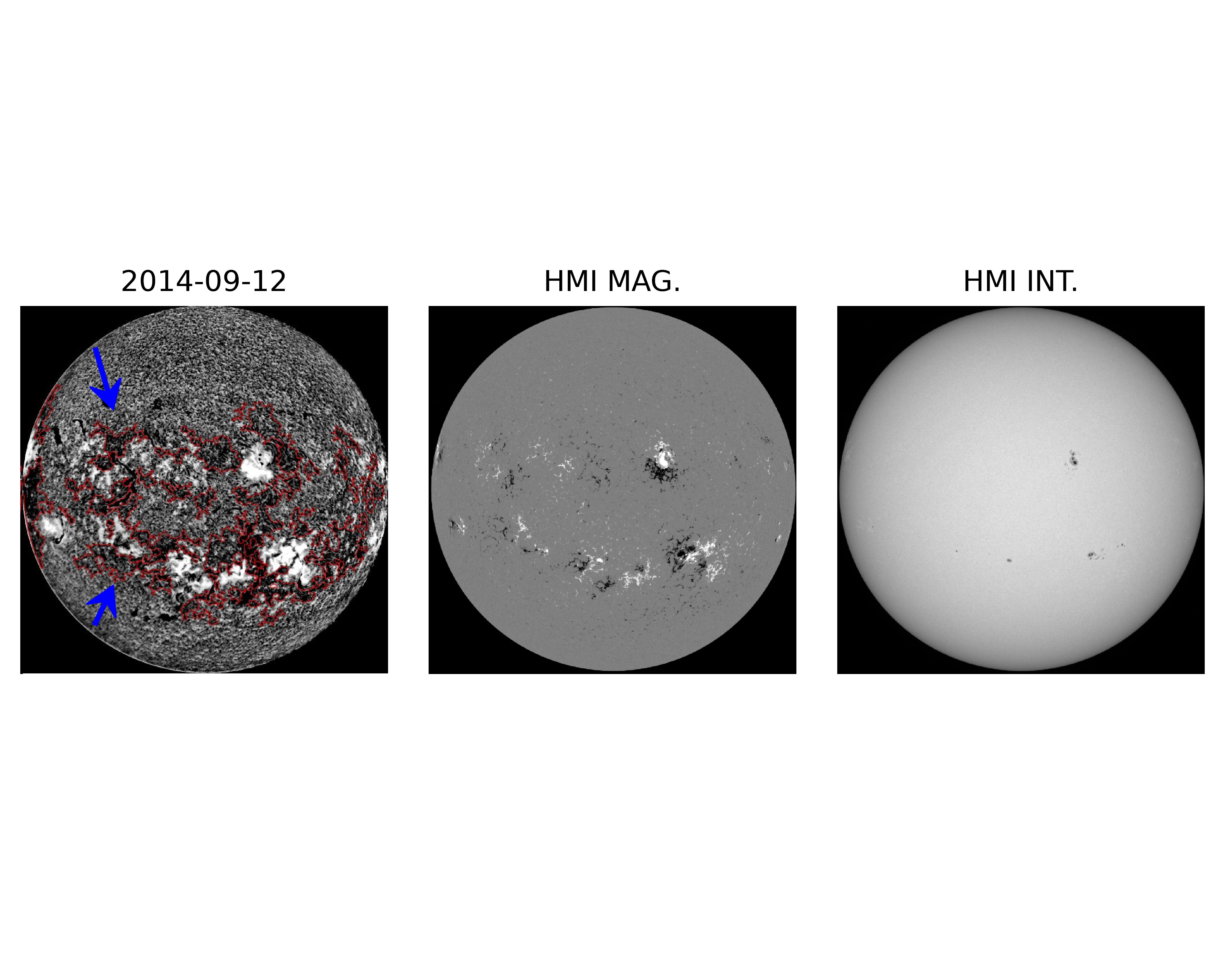}
\caption{ Left: ChroTel H$\alpha$ image showing the boundaries of circumfacular regions outlined by red contours. Blue arrows highlight two representative examples of circumfacular regions surrounding solar plages, though several others are visible. Center and Right: HMI magnetogram and continuum images acquired co-temporaneously with the ChroTel observation.} 
\label{fig:activenetwork}
\end{figure*}

\begin{figure*}
\centering
\includegraphics[width=0.93\textwidth,trim={2.5cm 35cm 2.cm 26cm},clip]{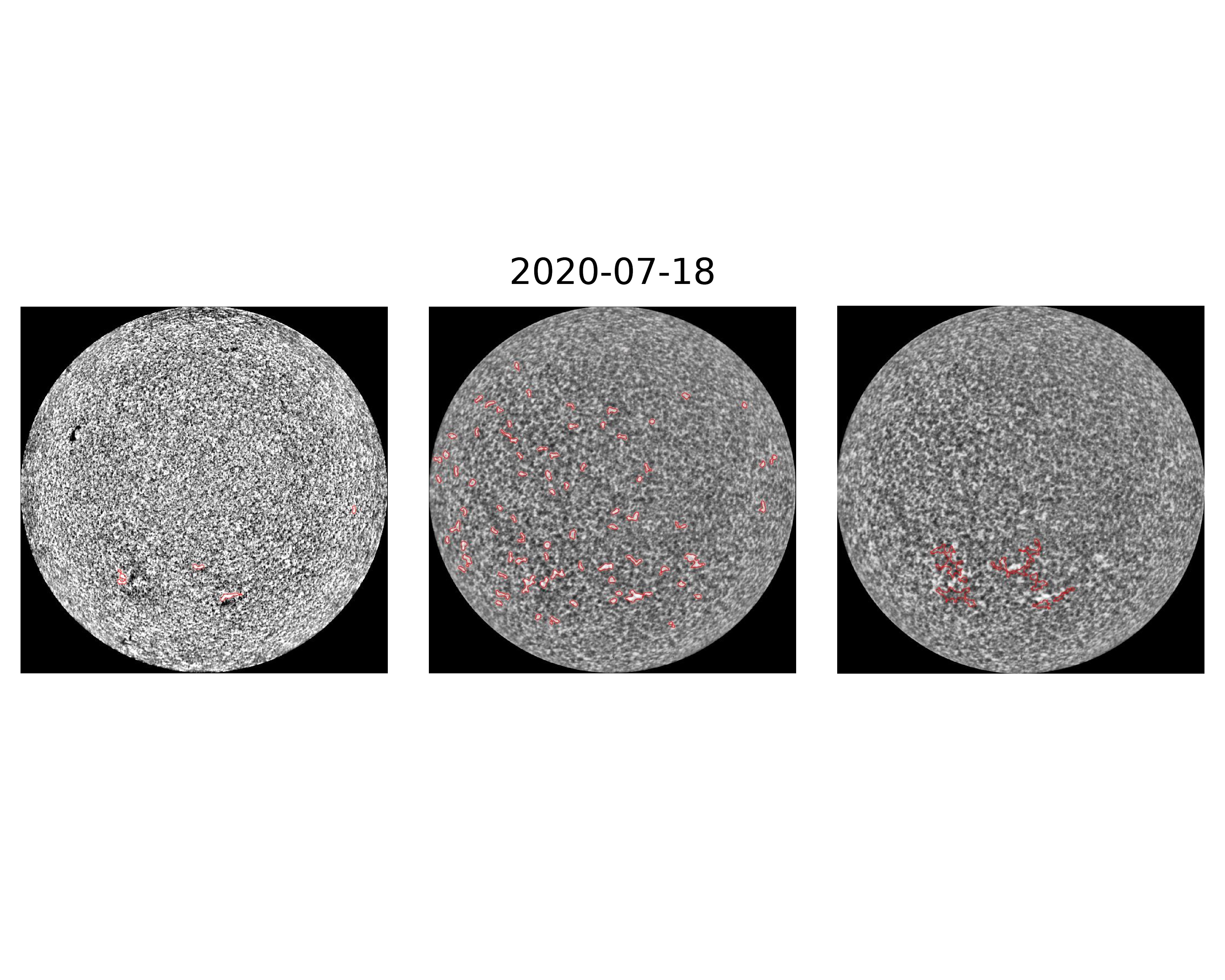}
\caption{Example of observations obtained during a period of low solar magnetic activity. Left: H$\alpha$ image with the contours of the identified excess regions overlaid in red. Center: \ion{Ca}{2}~K image with the contours of the identified excess regions overlaid in red. Right: \ion{Ca}{2}~K image with the contours of the identified circumfacular regions overlaid in red.
 } 
\label{fig:ims_at_min}
\end{figure*}

Figure~\ref{fig:AR11843} shows the evolution of a small, isolated, active region (AR11843) as it crosses the solar disk. The associated plage first becomes visible on 2013 September 17. By the following day, dark fibrillar structures have developed around the plage, marking the onset of the circumfacular region. On 2013 September 19, the circumfacular region is fully developed, and the identification algorithm detects it as a compact, approximately circular structure surrounding the active region. As the active region evolves, the circumfacular region gradually fragments and decreases in extent. By 2013 September 22, the active region is approaching the solar limb, and although the plage remains visible, the circumfacular region can no longer be identified. It should be noted, however, that circumfacular regions are frequently observed surrounding magnetic structures that are not associated with an active region. An example is shown in Fig.~\ref{fig:activenetwork}, where several circumfacular regions surround small plages in the absence of sunspots, as confirmed by the HMI continuum image. The corresponding magnetogram reveals diffuse magnetic concentrations of both polarities, most likely produced by the decay of large active regions during the previous solar rotation. We also identified numerous examples, particularly during solar minimum, of circumfacular regions associated with small unipolar and bipolar magnetic regions (ephemeral regions) that never developed sunspots. In many of these cases, the associated H$\alpha$ plage is faint or entirely absent, whereas the corresponding brightening is clearly visible in \ion{Ca}{2}~K observations, as shown for instance in Fig.~\ref{fig:ims_at_min}.

\subsection{Max-to-min variations} \label{sec:cycle}
\subsubsection{H$\alpha$ filtergrams}\label{Sec:Hafiltergrams}
\begin{figure*}
\centering
\includegraphics[width=0.8\textwidth]{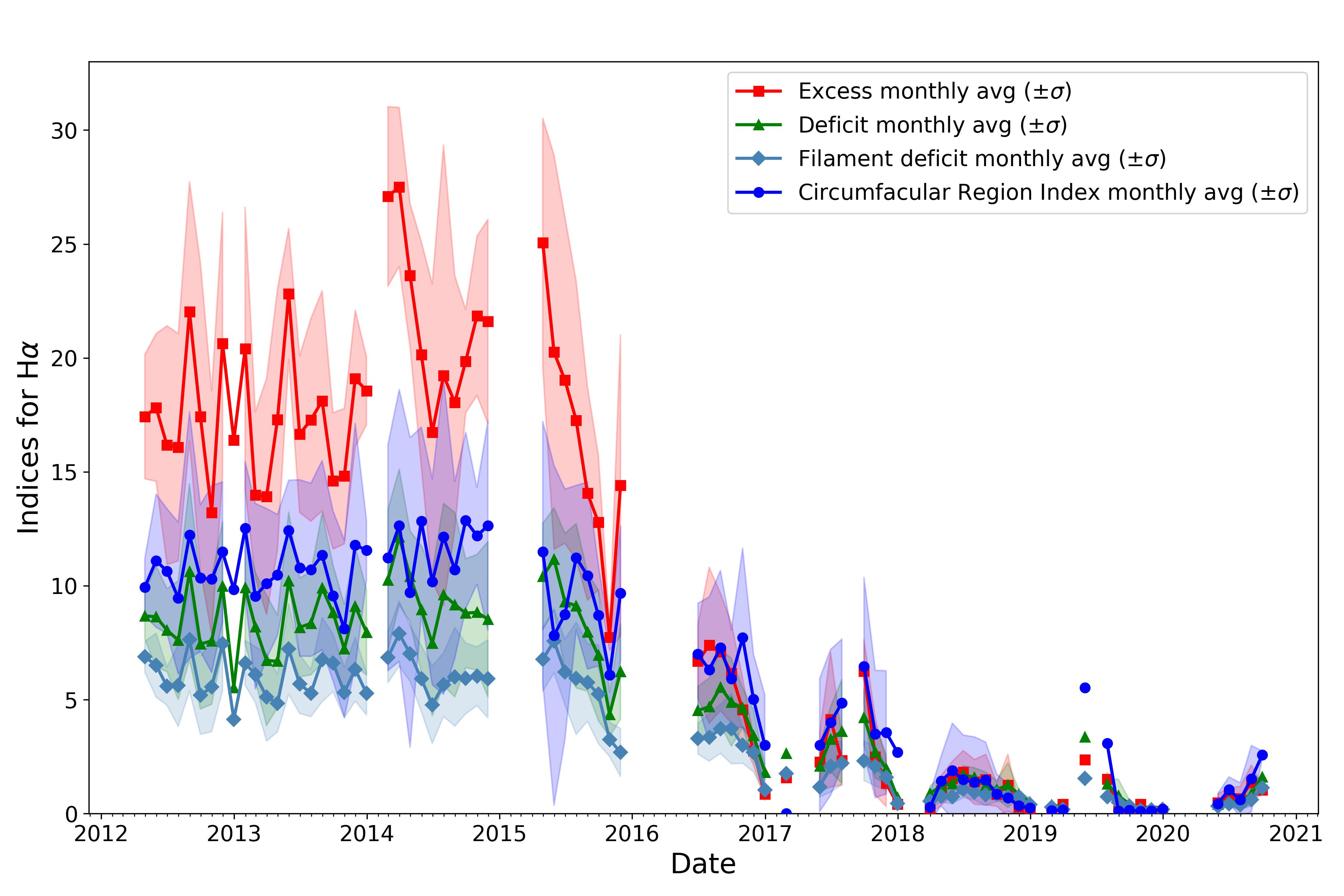}
\includegraphics[width=0.9\textwidth]{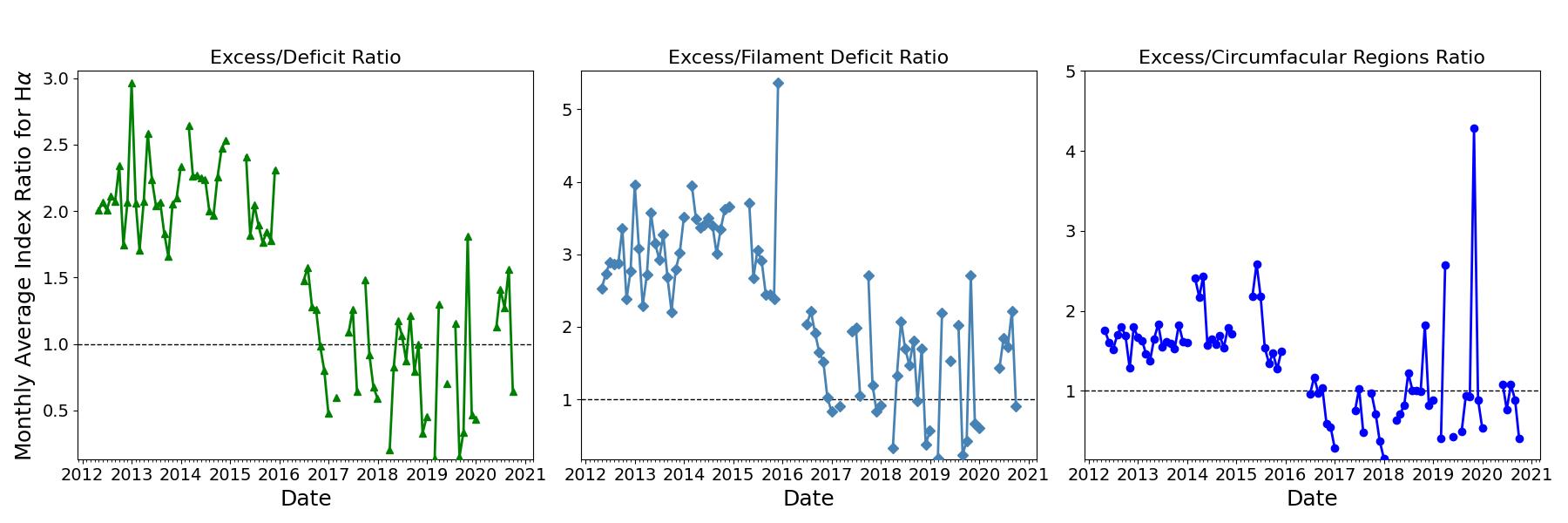}
\caption{Top: Monthly average of excess, deficit, filament, and circumfacular region indices observed in H$\alpha$ for solar cycle 24. The shaded areas indicate the $1\sigma$ standard deviation of the measurements within each monthly bin. Bottom: Ratios of excess index to deficit, filament deficit and circumfacular regions indices shown in the top panel.} 
\label{fig:min_max_variation}
\end{figure*}

The monthly variations of the different indices computed on H$\alpha$ images during solar cycle 24 are illustrated in Fig.~\ref{fig:min_max_variation} (top panel). All indices show a clear, in-phase variation with the magnetic activity, that during this cycle peaked in 2014 \citep[e.g.][]{penza2023,kaplan2024,liu2025}. As noted in \citet{diercke2022}, the maximum of the H$\alpha$ excess occurs in 2014, at the time of magnetic field reversal in the Southern hemisphere.  For the H$\alpha$ deficit, we found that the maximum occurs in 2014, with a secondary peak in the second half of 2015. We note that \citet{diercke2022} reported for the deficit a peak value at the end of 2015. This small discrepancy is attributed to the more restrictive definition of the deficit regions adopted in this work, which in \citet{diercke2022} also included circumfacular regions. Indeed, the circumfacular region index peaks in 2015. 
The H$\alpha$ excess is the largest among the indices and shows the largest absolute max-to-min variation, followed by circumfacular region index and the deficit index.  The filament index shows the smallest max-to-min variation. Interestingly, in the period 2017-2020, the circumfacular region index exceeds the excess index. To better compare the temporal variation of the indices, in Fig.~\ref{fig:min_max_variation} (bottom panel) we show the ratio between the H$\alpha$ excess and the other indices. The plot shows that during the maximum of activity the excess is on average 3-4 times larger than the filament deficit index, and about 2, 1.5 times larger than the deficit and circumfacular regions indices, respectively. The plot also shows that, as noted above, the circumfacular index slightly exceeds the excess between 2017 and 2020; during the same period the deficit approximately compares to the excess, while the filament deficit index is on average about 1.5 times smaller. It is important to note that the trends discussed above are average trends, as Fig.~\ref{fig:min_max_variation} shows a large dispersion for all indices, with the exception of the filament index. The dispersion is large especially during the maximum, while it reduces during the minimum, with the exception of the circumfacular regions index, for which the dispersion remains high even during the minimum. 

\begin{figure*}
\centering
\includegraphics[width=0.85\textwidth]{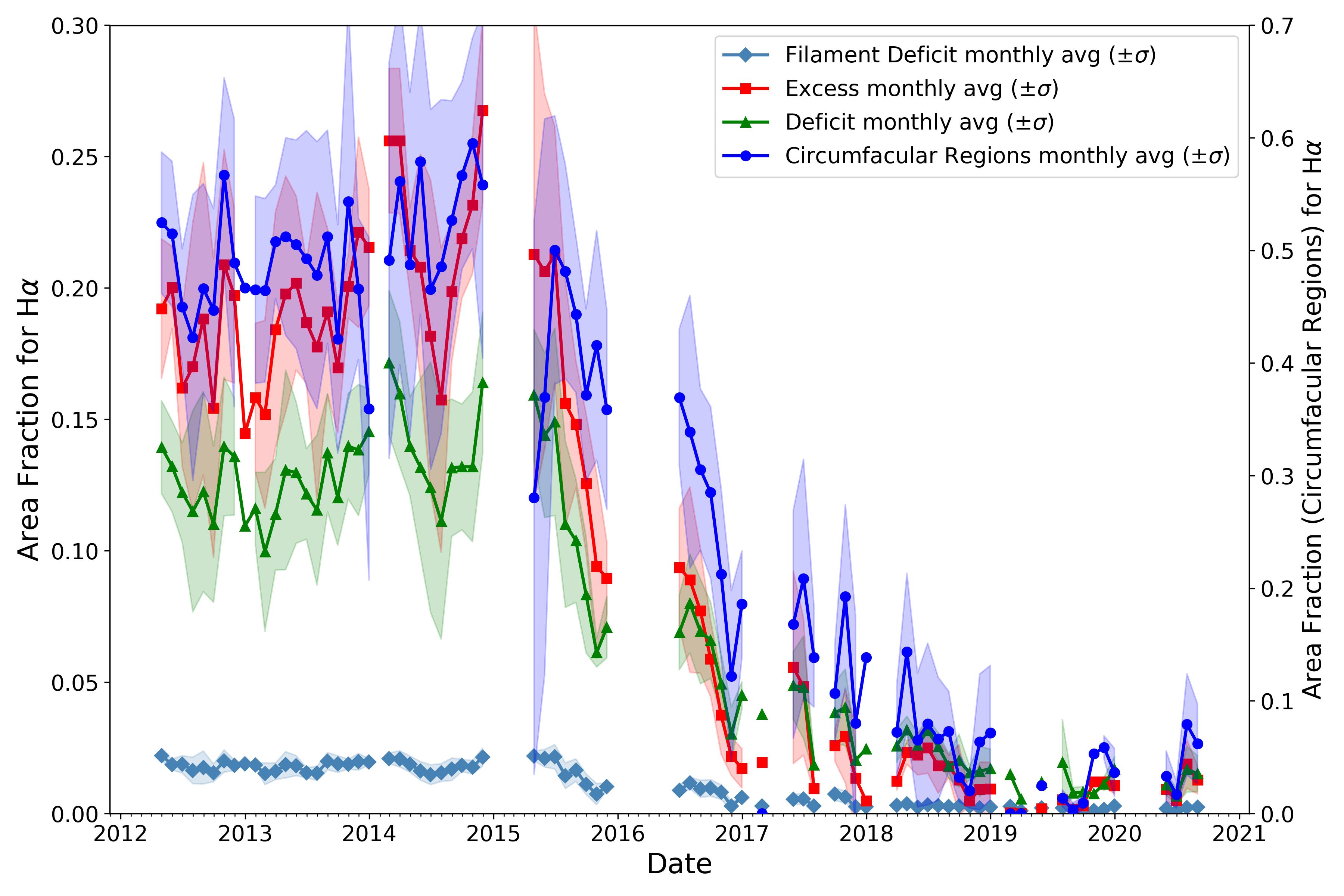}
\includegraphics[width=0.9\textwidth]{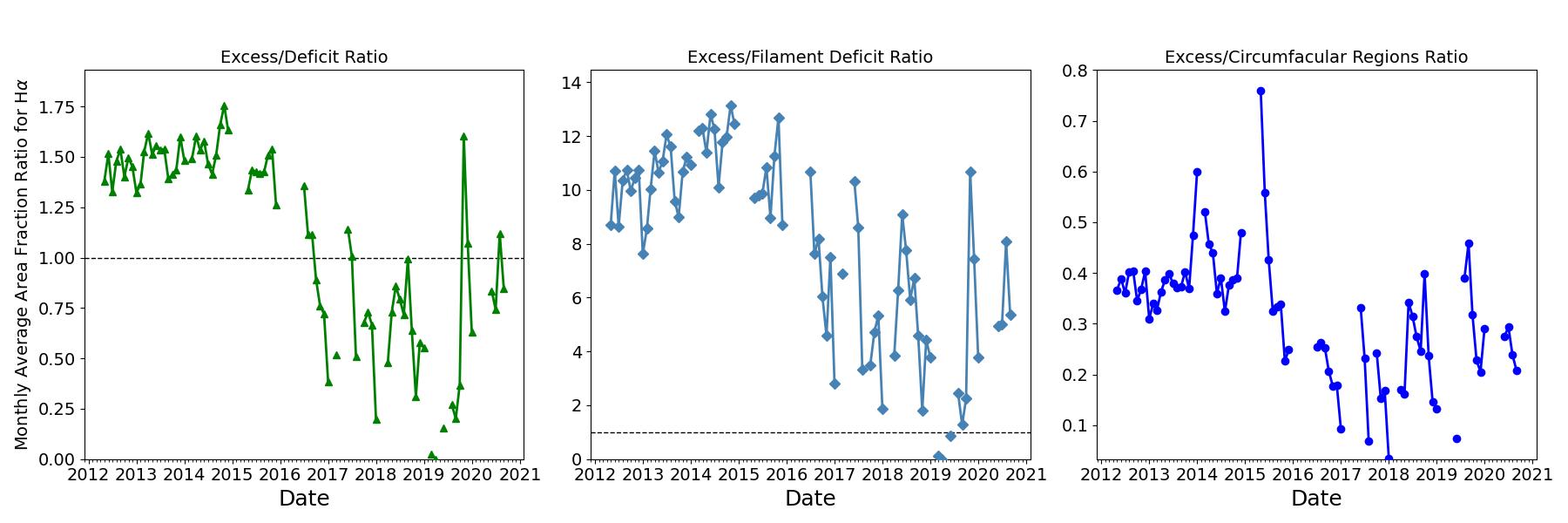}
\caption{Monthly average of excess, deficit and circumfacular regions area (in fraction of the solar disk) derived from H$\alpha$ observations during solar cycle~24. The y-axis on the right side is for the circumfacular regions. The shaded areas indicate the $1\sigma$ standard deviation of the measurements within each monthly bin. Bottom: Ratios of excess area to deficit, filament deficit and circumfacular regions areas shown in the top panel. } 
\label{fig:min_max_area_ratio}
\end{figure*}

To better understand the observed indices trends, we analyzed separately the temporal variations of the area fraction (computed relative to the area of the whole disk) and of the mean intensity of each structure. The plots in Fig.~\ref{fig:min_max_area_ratio} show the monthly averages of the area of each feature (top panel) and the ratio between the area of the excess regions and the area of other regions (bottom panels). Both plots show variations in-phase with the activity cycle, similar to the trends found for the indices (Fig.~\ref{fig:min_max_variation}).  Not surprisingly, circumfacular regions are the most extended structures, arriving to occupy more than 50\% of the solar disk during the maximum. The second most extended structures are excess regions, which can occupy more than 25\% of the solar disk. In comparison, deficit regions occupy even at maximum about 15\% of the disk, and filaments only a few percent. Interestingly, the excess regions area shows the fastest decrease right after the maximum, between 2015-2016. Plots in Fig.~\ref{fig:min_max_area_ratio} (bottom panel) show indeed a rapid decline of the area ratios at the beginning of the descending phase (especially visible for the excess/deficit area ratio), followed by a more gentle, albeit still with large fluctuation, decline during the minimum. In all cases, the minimum of the area ratios is reached in 2019, followed by an increase in 2020. In this respect, we note that, after reaching a minimum value close to zero during the minimum, the circumfacular regions index and area are higher than the other indices in 2020, thus suggesting that the circumfacular regions index may start rising more rapidly than other activity indicators. It would be interesting to cover a full cycle, to estimate the time in which, presumably during the rising phase, the excess exceeds the circumfacular regions index.  We also note that, in agreement with \citet{diercke2022}, both the dark and the filaments indices, together with their areas, drop right after November 2012 and March 2014, the times of the south and north magnetic field reversal, which \citet{diercke2022} ascribed to the disappearance of polar crown filaments. Those authors also noted an increase of the deficit in 2015, which they ascribed to the increase of dark features due to magnetic activity (i.e. filaments and sunspots). Similarly, we find  an increase of the deficit and filament deficit indices and in their respective areas, during the second half of 2015.    

\begin{figure*}
\centering
\includegraphics[width=0.9\textwidth]{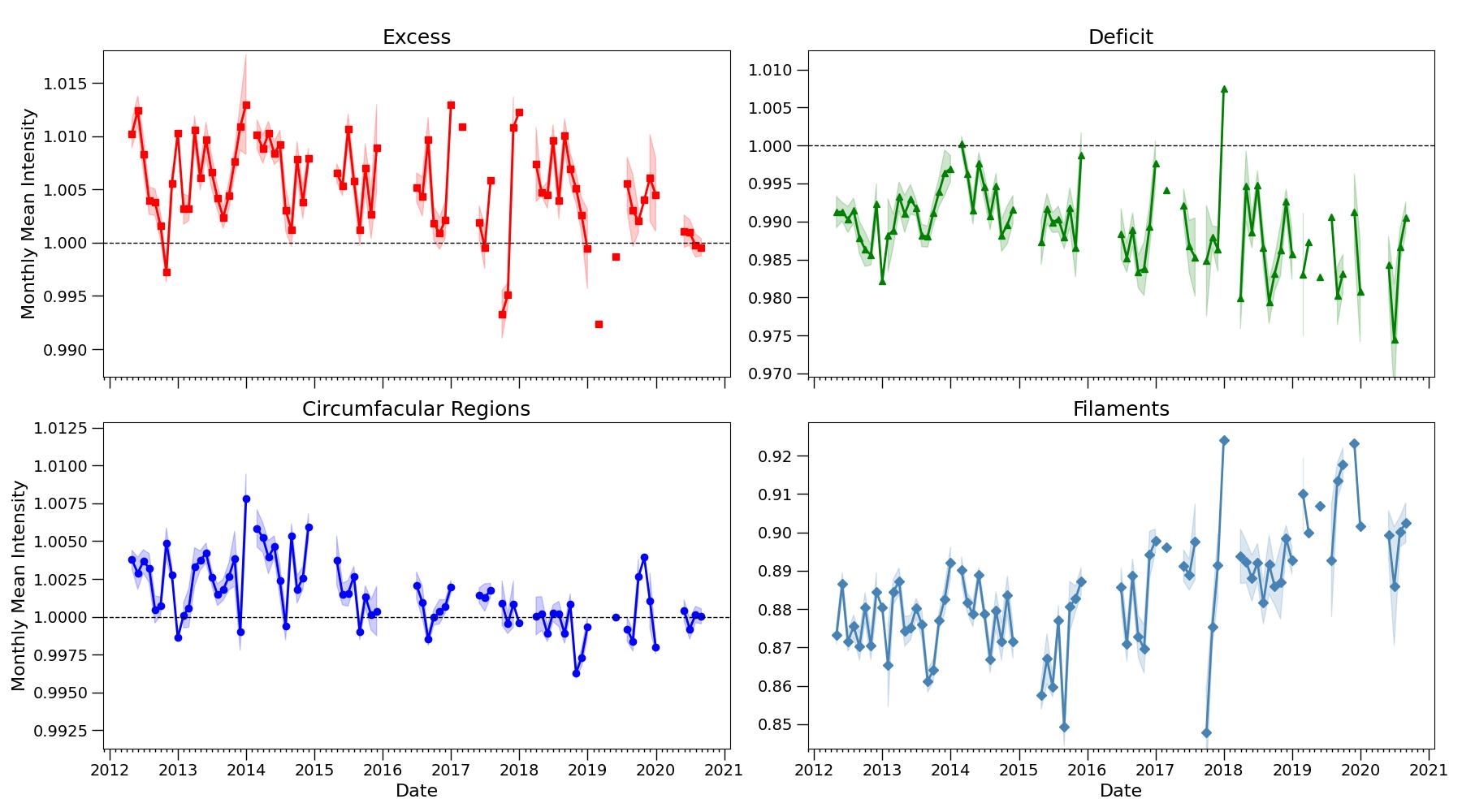}
\caption{Mean of H$\alpha$ intensity of excess, deficit, circumfacular regions and filaments. The shaded areas indicate the $1\sigma$ standard deviation of the measurements within each monthly bin.} 
\label{fig:mean_intensity}
\end{figure*}

Figure~\ref{fig:mean_intensity} shows the temporal evolution of the mean intensities of the different features, computed after normalizing each image by its median intensity, $I_{\mathrm{med}}$.
 The plots show that the average intensities of the excess and the circumfacular regions stay rather constant over the cycle, variations from maximum-to-minimum being on average smaller than 0.25\%. 
A somewhat larger variation of approximately 0.5\% is found for the deficit, while  the filament mean intensity shows a larger decrease of approximately 2.3\%. The excess, as expected, presents the largest mean intensity, followed by the circumfacular region mean intensity. As expected, both the deficit and the filament have intensities smaller than one, that is they are darker than the background. The small decrease of the mean intensities during the period of minima of both circumfacular regions and deficit should be ascribed to "contamination" from brighter structures into these regions, resulting mostly from the Opening procedures applied to the images as described in Sec.~\ref{sec:detection}. Concerning the circumfacular regions, inspection of the masks shows that these areas are interwoven with brighter filamentary structures. More specifically, they are largely composed of fibrils in which dark and bright features alternate (see Fig.~\ref{fig:index} and Fig.~\ref{fig:AR11843}). Contamination from nearby plage, the classification of the darkest pixels as deficit regions (see Sec.~\ref{sec:detection}), and the presence of comparatively bright fibrils all contribute to a net positive average intensity. This occurs despite the fact that, in direct visual inspection, circumfacular regions appear darker than their surroundings. 

The small max-to-min variations of the mean intensities, as opposed to the much larger variations found for the areas, indicate that indices variability is largely modulated by the area coverage of each feature, rather than their specific radiative emission.

\subsubsection{\ion{Ca}{2}~K filtergrams}
\begin{figure*}
\centering
\includegraphics[width=0.8\textwidth]{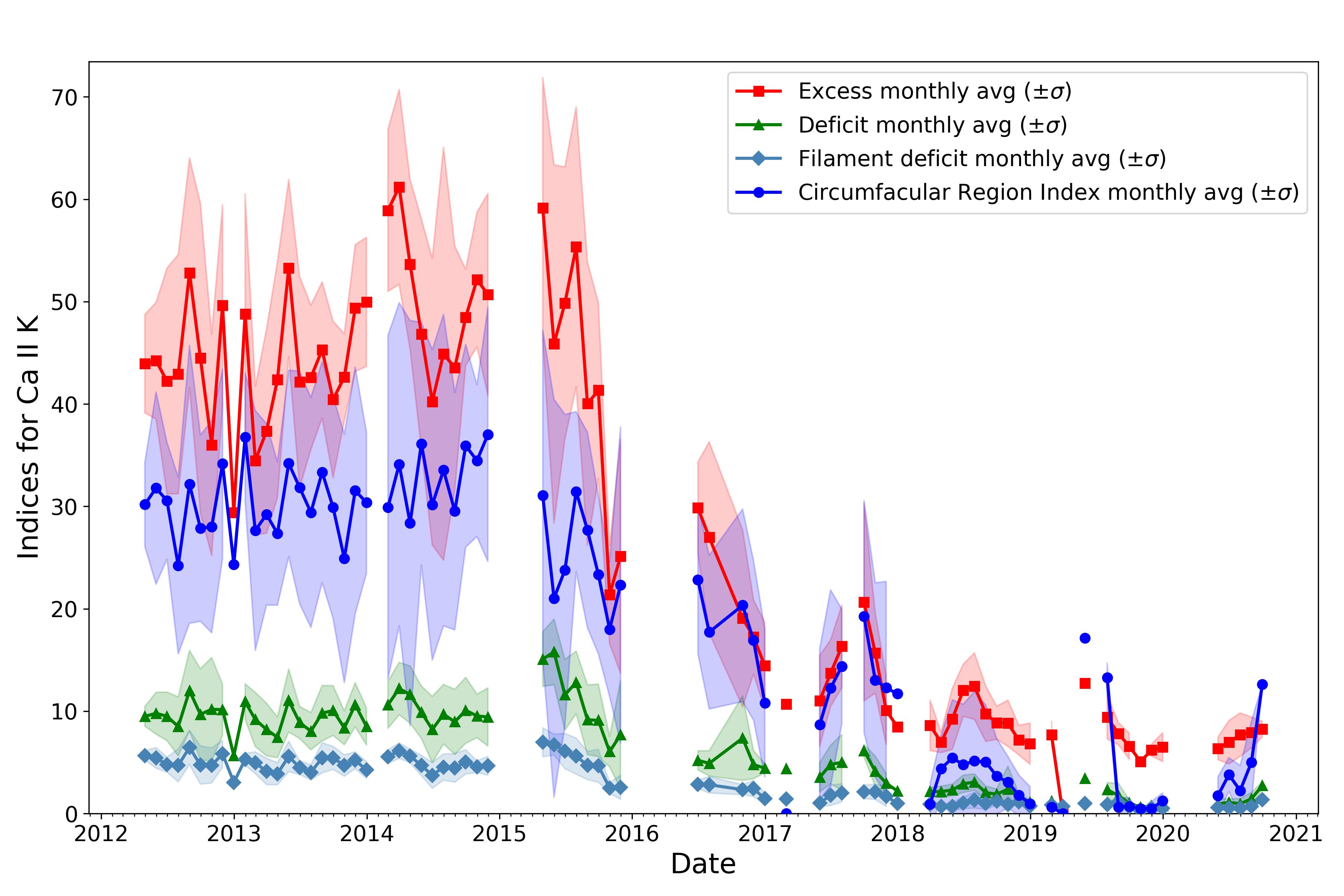}
\includegraphics[width=0.9\textwidth]{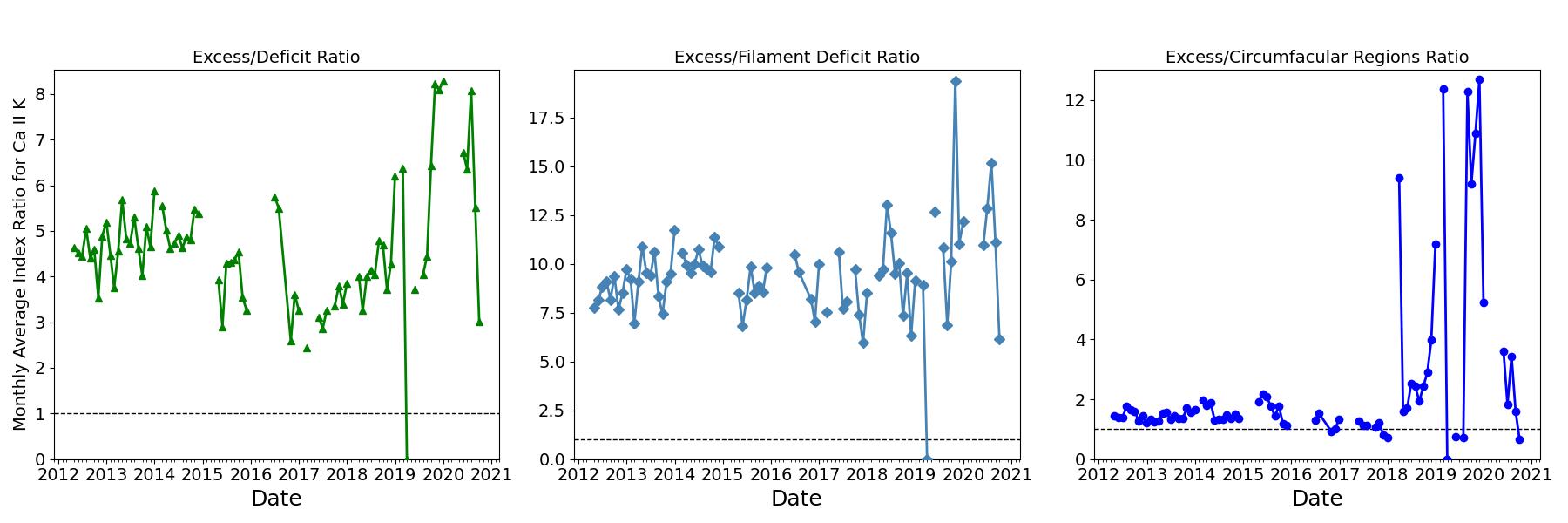}
\caption{Top: Monthly average of excess, deficit, filament, and circumfacular region indices observed in \ion{Ca}{2}~K for solar cycle 24. The shaded areas indicate the $1\sigma$ standard deviation of the measurements within each monthly bin. Bottom: Ratios of excess index to deficit, filament deficit and circumfacular region indices shown in the top panel.} 
\label{fig:min_max_variation_ca}
\end{figure*}

Figure ~\ref{fig:min_max_variation_ca} shows the monthly variations of the different indices estimated on \ion{Ca}{2}~K images (top panel), and the ratio of the excess to the other indices (bottom panel). Like  the H$\alpha$ indices, \ion{Ca}{2}~K indices show a clear modulation with the solar cycle.  

More specifically, inspection of Fig.~\ref{fig:min_max_variation_ca} shows that the excess has the largest value among the indices, and it presents two maxima of similar amplitudes in 2014 and in 2015. The excess index decreases rapidly at the end of 2015, but, unlike H$\alpha$, it remains larger than the other indices even during the minimum. The circumfacular regions index peaks  at the end of 2014, as observed in H$\alpha$, while both the deficit and the filament indices peak in 2015. Like for H$\alpha$, dimming of the deficit and filament deficit are observed at the end of 2012, in correspondence of the south magnetic field reversal, and at the beginning of 2014, in correspondence of the north magnetic field reversal. 

\begin{figure*}
\centering
\includegraphics[width=0.8\textwidth]{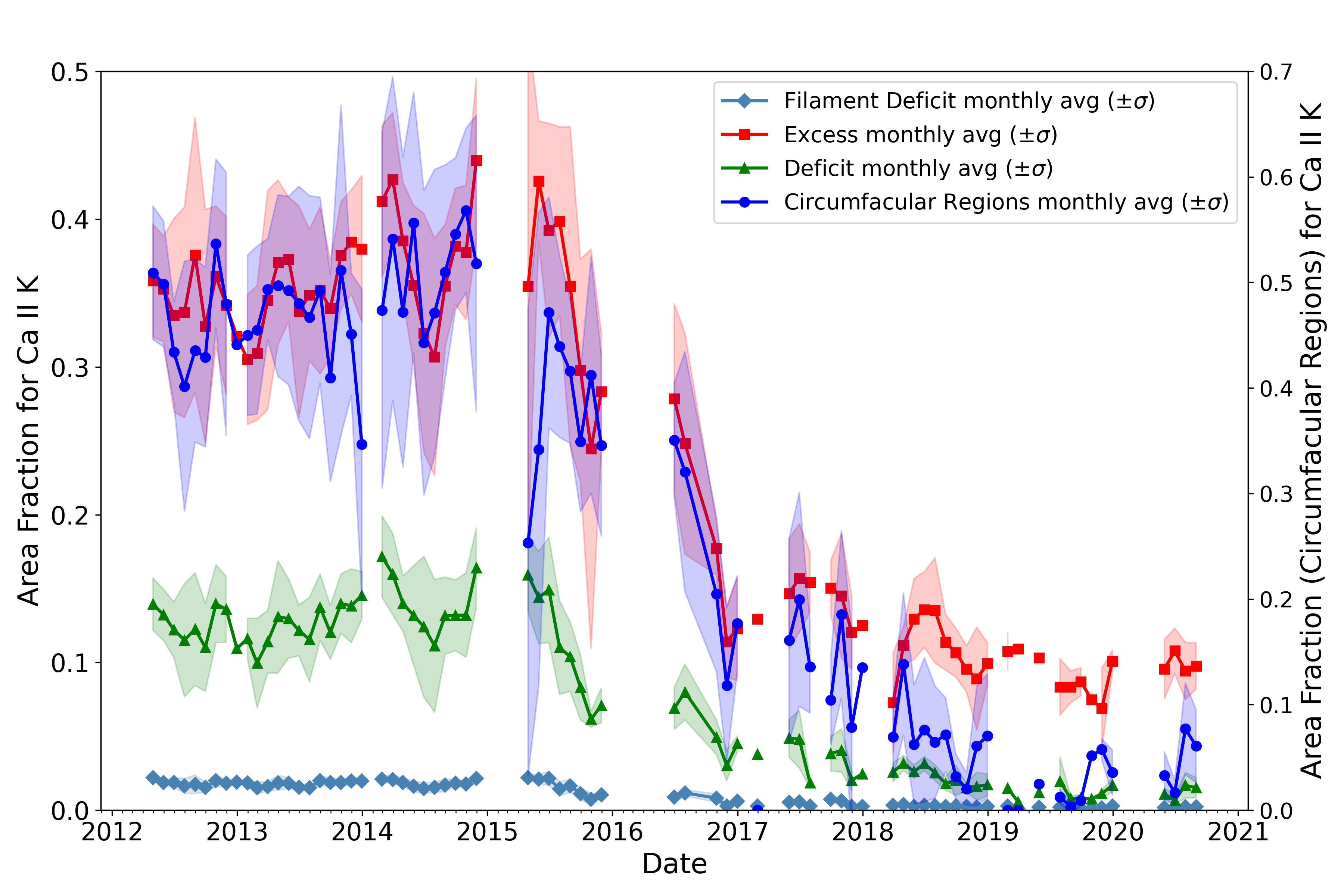}
\includegraphics[width=0.9\textwidth]{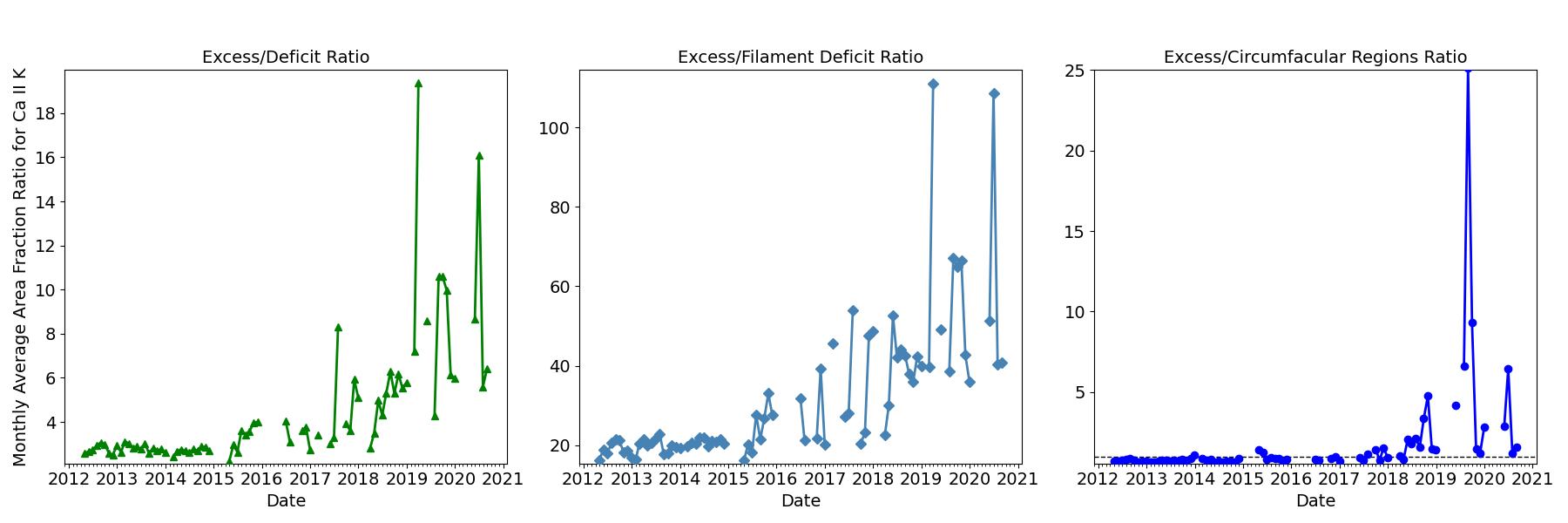}
\caption{Top: Monthly averages of excess,  deficit, filaments and circumfacular regions areas (expressed in fraction of the solar disk) during solar cycle 24. The shaded areas indicate the $1\sigma$ standard deviation of the measurements within each monthly bin. The excess area was derived from \ion{Ca}{2}~K observations, while the area deficit, filament and circumfacular regions were estimated from H$\alpha$ observations. Bottom: Ratios of excess area to deficit, filament deficit and circumfacular regions areas shown in the top panel.  }
\label{fig:excess_ratio_area_ca}
\end{figure*}

Figure~\ref{fig:excess_ratio_area_ca} shows the variation of the area of excess regions. For comparison the plot also shows variation of the area of the other features, already reported in Fig.~\ref{fig:min_max_area_ratio}. Like in H$\alpha$, the most abundant structures are circumfacular regions 
(which, as noticed in Sec.~\ref{Sec:Hafiltergrams}, cover more than 50\% of the disk), followed by the excess, which cover up to 40\% of the solar surface, the deficit and the filaments. Like for H$\alpha$, excess regions identified on \ion{Ca}{2}~K images show a large decrease during the descending phase.  
However, unlike in H$\alpha$ observations, both the ratio of indices (Fig.~\ref{fig:min_max_variation_ca}) and the ratio of areas (Fig.~\ref{fig:excess_ratio_area_ca}) increase toward the minimum. We will discuss such differences in detail in Sec.~\ref{sec:discussion}. 

\begin{figure*}
\centering
\includegraphics[width=0.9\textwidth]{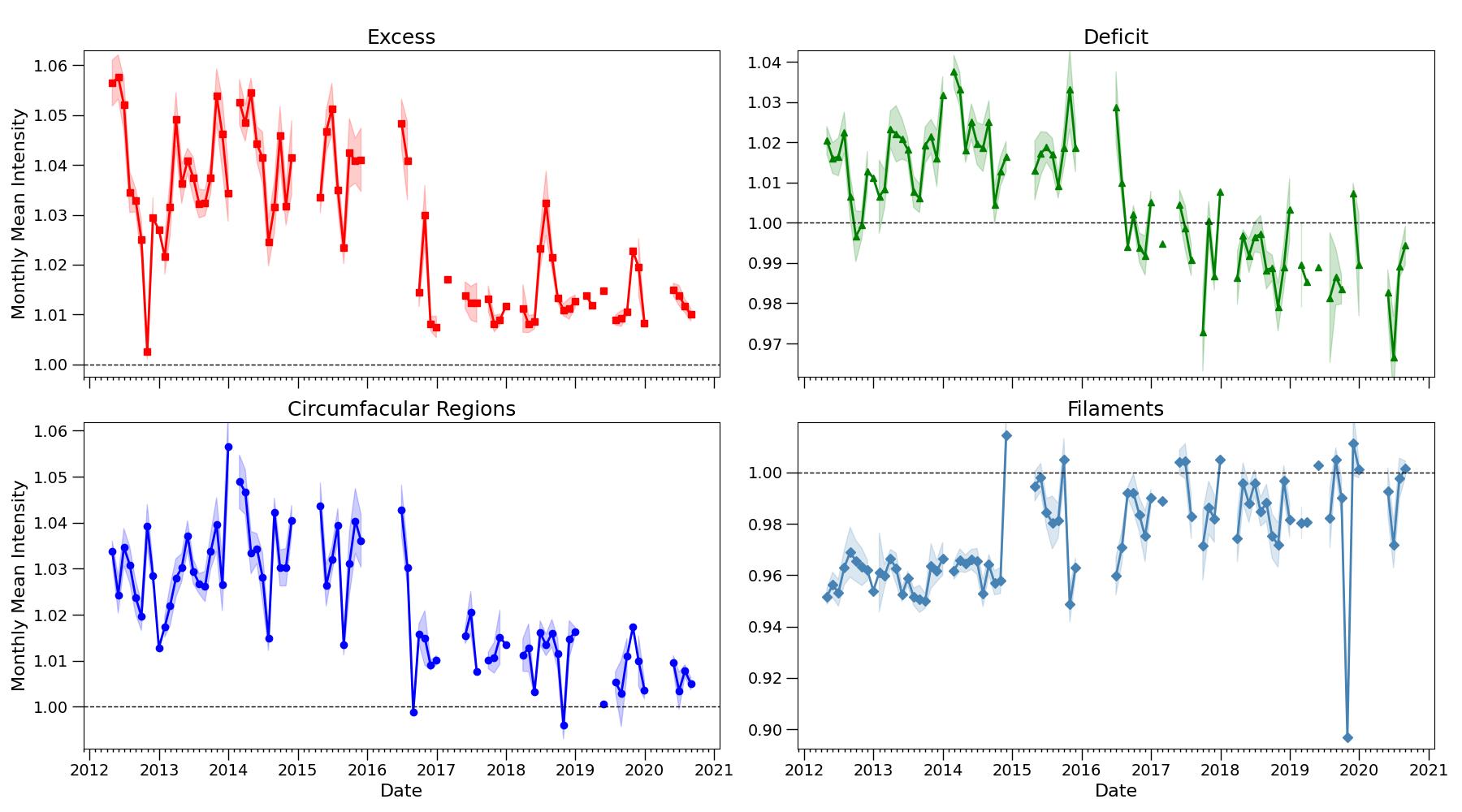}
\caption{Monthly averages of mean intensities of excess, deficit, circumfacular regions and filaments as derived from \ion{Ca}{2}~K observations. The shaded areas indicate the $1\sigma$ standard deviation of the measurements within each monthly bin.} 
\label{fig:ca_mean_intensity}
\end{figure*}

Over the solar cycle temporal scale, the mean intensities of all indices present small variations, as shown in Fig. \ref{fig:ca_mean_intensity}. Specifically, the maximum-to-minimum variations for the excess, deficit and circumfacular regions are 3\%, 4\% and 2\%, respectively, while the filament mean intensity varies by approximately 5\%. Because the decline of excess and circumfacular regions mean intensities is comparable, we conclude that the less steep decline of the circumfacular regions index respect to the excess index during the descending phase is due to the less steep decline of the circumfacular regions area coverage. The excess presents the largest mean intensity at all times, followed by the circumfacular regions.  For these features, the average intensities are always larger than unity, meaning that they are brighter than the background. The deficit presents mean intensity larger than one during the maximum, but smaller than one (meaning these structures are darker than the background) during the descending phase and the minimum. Opposite trend is found for the mean intensity of the filaments, which are darker at the maximum rather than at the minimum.


\subsection{Carrington rotation: Active Longitudes} \label{sec:activelong}

\begin{figure*}
\centering
\includegraphics[width=0.9\textwidth]{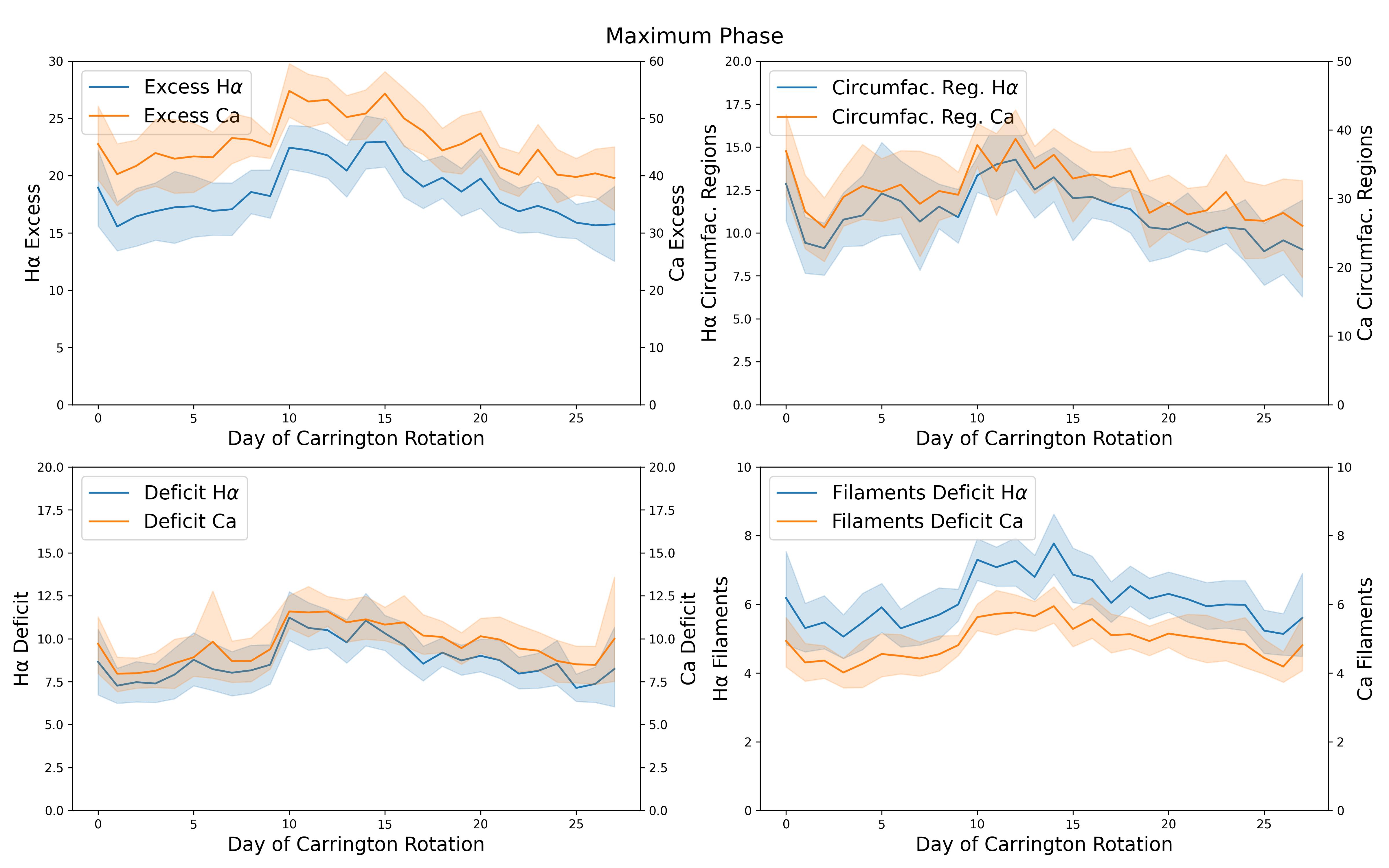}
\includegraphics[width=0.9\textwidth]{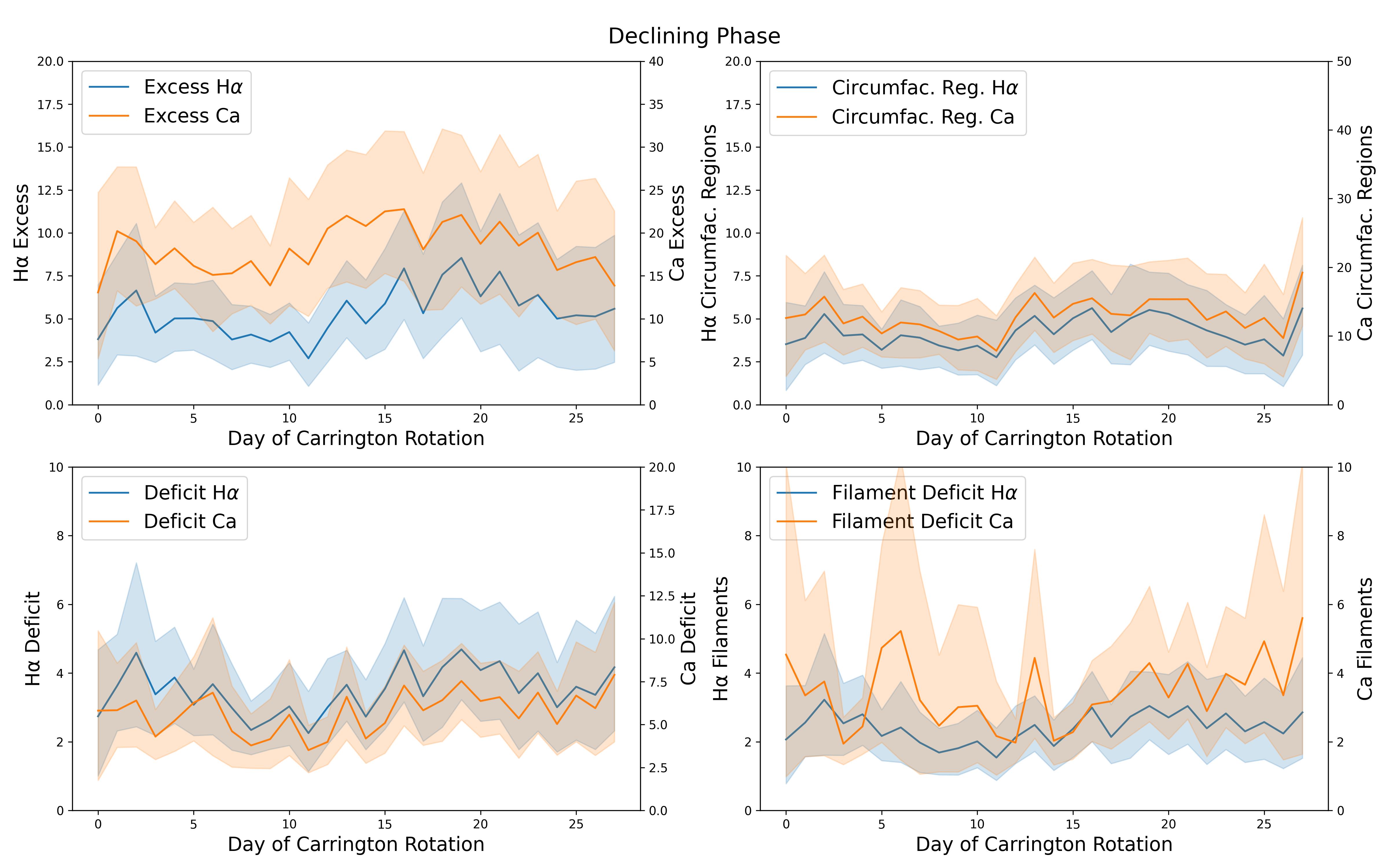}
\caption{Active Longitude for H$\alpha$ and \ion{Ca}{2}~K for excess, circumfacular region index, and deficit for all dark regions and for filaments (from left to right and top to bottom). Continuous lines and shaded areas represent the averages and the standard deviations, respectively, computed each Carrington rotation day. Top: data acquired during the maximum phase. Bottom: data acquired during the declining phase.} 
\label{fig:active_longitude_maxandmin}
\end{figure*}

Following \citet{diercke2022}, we investigated the variability on the rotational scale \citep[i.e. 27.2753 days,][]{knaack2004} for all Carrington rotations.  ChroTel observed a total of 84 Carrington rotations, 39 in the maximum (2012-2014) and 45 in the descending phase and minimum (2015-2020). 
Images were grouped according to their Carrington evolution and the average and standard deviation of the different indices were computed for each group (each Carrington day). Figure~\ref{fig:active_longitude_maxandmin} shows results obtained during the maximum phase  and the descending phase of solar cycle 24. 

In agreement with \citet{diercke2022} we found that during the maximum phase the H$\alpha$ excess and the H$\alpha$ deficit show a bell-shape trend, with two peaks  approximately between 10 and 15 days, or active longitudes between approximately 150-180 $\deg$ in the Carrington reference frame. Interestingly, a similar trend is found also for the circumfacular regions index, although the bell-shape trend is less pronounced. During the descending phase the bell-shape vanishes for all three indices, while various other peaks are clearly visible. The bell shape is less pronounced for the circumfacular regions index because these regions are typically much larger and extend over different longitudes, so that the association to a specific location over the disk is less accurate.
Comparing the two figures, we also find that both the \ion{Ca}{2}~K and H$\alpha$ excess indices are three times larger during the period of maximum than in the declining phase. Similarly, on average the other indices show a decline between two and three times from the maximum to the minimum, the changes being closer to three times for the indices derived from \ion{Ca}{2}~K data, most likely due to the larger change in intensities from maximum-to-minimum found for this waveband.

The plots show also that all indices derived from \ion{Ca}{2}~K images closely follow the trends of indices derived from H$\alpha$, in both the maximum and declining phase.

\section{Discussion}\label{sec:discussion}

\subsection{Comparison between H$\alpha$ and \ion{Ca}{2}~K indices}


The indices derived from H$\alpha$ and \ion{Ca}{2}~K overall exhibit similar temporal behavior on both decadal and Carrington time scales. Table~\ref{tab:correls} lists the Pearson and Spearman correlation coefficients between the \ion{Ca}{2}~K and H$\alpha$ indices, evaluated at activity maximum and during the descending/minimum phase using Carrington-rotation–averaged data. In all cases, the correlations are positive and statistically significant at the 1\% level, with the exception of the filament indices during the declining phase. The strongest correlations are observed for all features during activity maximum, while the correlation coefficients decrease  toward solar minimum. The weakest correlations are found for filaments during the declining phase; in this case, the corresponding p-value exceeds the 1\% significance threshold, indicating that the inferred correlation for these features is not statistically robust.
For completeness, we note that for deficit, filament, and circumfacular regions the same H$\alpha$-derived masks are applied to both H$\alpha$ and \ion{Ca}{2}~K images. Because the variability of the corresponding indices is strongly driven by changes in the area of these masks, this common spatial definition can contribute to the high correlations measured between the two passbands.

 However, when comparing the results obtained in the two wavebands several differences also emerge. The excess index is larger, approximately two to three times higher in \ion{Ca}{2}~K than in H$\alpha$. 
 The circumfacular regions index is also enhanced in \ion{Ca}{2}~K, by a factor of approximately three. In contrast, the deficit index shows similar values in both spectral regions, while the filament deficit index remains broadly comparable as well, with H$\alpha$ displaying slightly higher values due to the lower mean intensity of filaments in this spectral region (see Fig.~\ref{fig:mean_intensity} and Fig.~\ref{fig:ca_mean_intensity}). 

Concerning the trends over the decadal scale, all indices show a general positive correlation with the activity cycle, being larger during the maximum of solar cycle 24 and smaller during the minimum period between solar cycles 24 and 25. However, the deficit index in H$\alpha$ shows a major peak in 2014, followed by a second peak 2015, while in \ion{Ca}{2}~K a single large peak is found in 2015. At both wavelengths, during the descending phase all indices decrease more gradually than the excess index. However, in H$\alpha$ the excess index is often zero (or close to) during the minimum, while it remains elevated (at roughly a fifth of the value found during the maximum) in \ion{Ca}{2}~K. The relatively high value of the excess index in \ion{Ca}{2}~K respect to H$\alpha$ found at the minimum explains the different trends of the indices ratios found for the two wavebands (bottom panels of Fig.~\ref{fig:min_max_variation} and Fig.~\ref{fig:min_max_variation_ca}), which are clearly in phase with the activity for H$\alpha$ and in counterphase (especially for the excess/circumfacular regions ratio) in \ion{Ca}{2}~K. These differences are due to the fact that during the minimum in H$\alpha$ the fraction of the solar disk covered by the excess regions is less than a few percent (Fig.~\ref{fig:min_max_area_ratio}), while in \ion{Ca}{2}~K the area coverage is above 10\% (Fig.~\ref{fig:excess_ratio_area_ca}).

    
\begin{table}
\centering
\caption{Correlations between H$\alpha$ and \ion{Ca}{2}~K indices averaged by Carrington rotation.}
\label{tab:correls}
\begin{tabular}{lcccc}
\hline\hline
 & \multicolumn{2}{c}{Maximum} &
 \multicolumn{2}{c}{Declining/Minimum} \\
\cline{2-3}\cline{4-5}
Index & Spearman & Pearson & Spearman & Pearson \\
\hline
Excess             & 0.94 & 0.91 & 0.71 & 0.70 \\
Deficit            & 0.92 & 0.91 & 0.84 & 0.81 \\
Filament   & 0.96 & 0.96 & 0.54 & 0.51 \\
Circum.              & 0.92 & 0.92 & 0.90 & 0.91 \\
\hline
\end{tabular}
\end{table}
At both wavelengths the excess regions exhibit the highest mean intensity values. Nonetheless, several distinctions emerge: for every type of structure, the mean intensity measured in \ion{Ca}{2}~K is higher than in H$\alpha$; in \ion{Ca}{2}~K  the deficit regions are, on average, slightly brighter than the background up to 2017, becoming darker during solar minimum; the mean intensity of filaments displays the same trend of increasing from maximum to minimum activity,  but is higher (filaments are less dark) by approximately 10\% in \ion{Ca}{2}~K.

Concerning the circumfacular regions, their mean intensity  is positive in both \ion{Ca}{2}~K and H$\alpha$.   
However, for the \ion{Ca}{2}~K the positive values arise not only from the image-opening procedure described earlier but also from the intrinsic brightness distribution within the circumfacular areas themselves: bright structures appear brighter, while dark fibrils appear less dark, leading to an overall positive average intensity.

Finally, it should be noticed that for deficit, filament, and circumfacular regions, the masks derived from H$\alpha$ observations were also applied to the corresponding \ion{Ca}{2}~K images. Therefore, the quantities derived from \ion{Ca}{2}~K for these
structures should not be interpreted as measurements of independently identified \ion{Ca}{2}~K features. Rather, they quantify the \ion{Ca}{2}~K emission originating from the locations occupied by the corresponding H$\alpha$ structures. This distinction is particularly important for circumfacular regions, since \citet{bumba1965} reported that circumfaculae identified in \ion{Ca}{2}~K are generally smaller than their H$\alpha$ counterparts. Consequently, the H$\alpha$-defined circumfacular masks may include pixels that would not be classified as circumfacular in \ion{Ca}{2}~K, and that most likely contribute increasing their average intensity.

\subsection{Comparison with results in \citet{diercke2022}}\label{sec:comp_diercke}
The analysis presented in this study builds upon the work of \citet{diercke2022}, who analyzed the same ChroTel H$\alpha$ dataset and performed a similar investigation of excess and deficit regions. A comparison of the two studies reveals several similarities. In particular, the temporal evolution of the excess and deficit indices, both over the solar cycle and on Carrington-rotation timescales, is largely consistent, with maxima occurring at the same epochs. The absolute values of the indices, however, differ significantly, with those derived in the present study being systematically larger. For example, we find monthly peak values of approximately 27 and 12 for the excess and deficit indices, respectively, whereas \citet{diercke2022} reported peak values of about 17 and 3.5.

These differences should be primarily ascribed to the different feature-identification procedure adopted in this work. As discussed in Section~\ref{sec:detection}, our method identifies larger excess and deficit regions than the algorithm of \citet{diercke2022}. Specifically, the area of excess regions is up to five times larger, while the area of deficit regions is up to eight times larger than those reported in the previous study. Consequently, because a larger number of pixels with intensities closer to the quiet-Sun level are included, the average intensities of both excess and deficit regions are lower (i.e., closer to unity) than those reported by \citet{diercke2022}.

The comparison with the results of \citet{diercke2022} also serves to illustrate the sensitivity of the derived areas and index values to the adopted feature-identification method. The substantially different areas and index amplitudes obtained in the two studies demonstrate that these quantities are method-dependent. Nevertheless, both analyses recover similar overall temporal behavior of the excess and deficit indices and their modulation over the solar cycle, although some differences are found in the timing and relative amplitudes of individual maxima. This agreement suggests that, despite their dependence on the segmentation method, the indices provide robust tracers of the overall evolution of magnetic activity. 

\subsection{Implications for spectral indices variability}

Solar indices are measures derived from observations of specific solar features or emissions, such as sunspots, plages, faculae, chromospheric lines, or radio flux, that track different aspects of solar magnetic activity and variability \citep[e.g.][]{ermolli2015,usoskin2023}. Common examples include the international sunspot number, the F10.7 cm radio flux, and chromospheric indices based on \ion{Ca}{2}~K or H$\alpha$ emission, each sampling activity at different heights in the solar atmosphere. These indices are widely used as proxies for solar irradiance and magnetic activity because they often extend over long time spans, are relatively easy to measure, and correlate well with more complex or less frequently available physical quantities, as for instance UV radiation \citep[e.g.][]{dudok2009}. As a result, solar indices are fundamental in studies of solar variability, space weather, and Sun–climate connections, enabling both empirical reconstructions and physics-based modeling of solar influences on the heliosphere and Earth’s environment \citep[e.g.][]{petrie2021}. However, different indices respond in different ways to magnetic activity, due to various factors, as their different sensitivity to the different magnetic structures, the formation height of the observational wavelengths from which they are derived, and their sensitivity to atmospheric physical perturbations produced by the magnetic activity. As a result, different indices may present complex variations with each other during the cycle, and may differ from cycle to cycle \citep[e.g.][]{criscuoli2016, mursula2024}. 

In Sec.~\ref{sec:intro} we mention the complex relation displayed by indices extracted from \ion{Ca}{2}~K and H$\alpha$ observations. Although those indices were mostly derived from spectroscopic observations, as for instance the core-to-wing ratio, line width and equivalent width of lines, while our analysis is performed using broad band images, results presented in this paper offer a clue to interpret the relation between \ion{Ca}{2}~K and H$\alpha$ spectroscopic observations.

First, we note that in H$\alpha$ the ratio between the excess and filament areas is always greater than unity, even during periods of low activity. While this ratio decreases as activity declines, whereas it increases in \ion{Ca}{2}~K, the fact that it remains consistently above one indicates that filaments are unlikely to produce an anti-correlation between the \ion{Ca}{2}~K and H$\alpha$ spectral indices, as the negative contribution of filaments to the line-core intensity is likely to be over-compensated by the positive contribution from excess regions (which broadly correspond to plage areas). This interpretation is supported by the increase in the mean filament intensity (i.e., filaments becoming less dark) during low-activity periods, whereas the mean intensity of excess regions remains approximately constant. In contrast, in H$\alpha$ the ratio of the excess area to the areas of deficit and circumfacular regions drops below unity at activity minimum for the former and remains below unity at all times for the latter. In both cases, the areas of deficit and circumfacular regions exceed that of the excess by factors of ten or more during low-activity periods, suggesting that these features dominate H$\alpha$ variability near solar minimum. The magnitude and sign of their contribution, however, depend on the specific spectral index considered. For the core-to-wing ratio, we speculate that the contribution of circumfacular regions is likely negative, since these regions correspond predominantly to quiet-Sun areas in photospheric observations, and their intensity in the core of H$\alpha$ is likely to be lower than the one of quiet Sun, as found in Ca II 854.2 nm spectroscopic observations (see Sec.~\ref{sec:intro}). 

The contribution of deficit regions to the H$\alpha$ core-to-wing ratio is less straightforward to infer, as this class includes a heterogeneous mixture of features such as filaments, sunspots, and the darkest portions of circumfacular regions, that affect the ratio in different ways. For example, \citet{criscuoli2023} showed that sunspots contribute positively to the core-to-wing ratio, despite having lower radiative emission than the quiet Sun in both the continuum and the line core.

In contrast, for \ion{Ca}{2}~K the ratio between the area of excess regions and that of all other regions exhibits the opposite behavior, increasing during periods of low activity. This suggests that the reduced correlation between the H$\alpha$ core-to-wing ratio and the \ion{Ca}{2}~K index is primarily driven by the disappearance of plage and network regions in H$\alpha$ observations. This interpretation is consistent with the conclusions presented in \citet{criscuoli2023}.

\section{Conclusions}

This paper presents the first comprehensive comparison of the photometric and geometric properties of circumfacular regions observed in H$\alpha$ and \ion{Ca}{2}~K since the seminal work of \citet{bumba1965}.  We also compare the properties of circumfacular regions with those of excess regions (which broadly correspond to plages), deficit regions, and filaments. The main results of our study, based on the analysis of ChroTel observations acquired between 2012 and 2020, are the following: 
\begin{itemize}
    \item In agreement with \citet{bumba1965}, we found that circumfacular regions are more difficult to detect in \ion{Ca}{2}~K than in H$\alpha$.
   \item Visual inspection of relatively isolated active regions suggests that circumfacular regions develop after the associated H$\alpha$ plage brightening and persist during the subsequent evolution of the active region, in qualitative agreement with \citet{bumba1965}. The limited temporal sampling of the observations does not allow us to establish a characteristic time delay. Moreover, circumfacular regions persist long after the disappearance of sunspots, and are detected in H$\alpha$ images even when plages are small and barely discernible.
    \item Following \citet{diercke2022}, we defined photometric indices for excess, deficit, circumfacular regions and filaments. \citet{diercke2022} pointed out that both the excess and deficit are good tracers of activity. Here we find that circumfacular regions and filament indices are also good tracers of magnetic activity. 
    \item Indices extracted from H$\alpha$ and \ion{Ca}{2}~K images show strong, positive correlation during different phases of the activity cycle, with the exception of the filament index during the descending/minimum phase, for which the correlation is not statistically significant, likely because of the reduced number of filament observations.
    \item Circumfacular regions are the most extended features  
    among the ones investigated in both wavebands during activity maximum. However, during the minimum phase of cycle 24 and beginning of cycle 25, their area is exceeded by the one of excess regions singled out in \ion{Ca}{2}~K images.
    \item The solar-cycle evolution of all indices is modulated primarily by changes in area coverage, whereas the mean intensity of the individual structures remains nearly constant. 
    \item Our results provide insight into the contribution of different chromospheric structures to disk-integrated spectral indices. In both H$\alpha$ and \ion{Ca}{2}~K, the ratio between the excess and filament deficit indices remains greater than unity throughout the solar cycle, suggesting that filaments alone are unlikely to produce the observed decrease in correlation between H$\alpha$ and \ion{Ca}{2}~K activity indices. In contrast, during the declining phase and solar minimum, the deficit and circumfacular region indices exceed the excess index in H$\alpha$ (mostly due to the almost disappearance of plage/network features in this spectral range), whereas the excess index remains dominant in \ion{Ca}{2}~K. These differences suggest that deficit and circumfacular regions are more likely to drive the reduced correlation between H$\alpha$ and \ion{Ca}{2}~K activity diagnostics during periods of low activity. Therefore, our results indicate that circumfacular regions should be considered alongside plages and filaments when interpreting chromospheric variability. Nevertheless, we stress that spectroscopic observations are required to confirm this interpretation because different chromospheric structures contribute differently to individual spectral diagnostics.

\end{itemize}

\begin{acknowledgments}
The National Solar Observatory (NSO) is operated by the Association of Universities for Research in Astronomy, Inc. (AURA), under cooperative agreement with the National Science Foundation.
ChroTel is operated by the Institute for Solar Physics (KIS), Freiburg, Germany, at the Spanish Observatorio del Teide, Tenerife, Canary Islands. The ChroTel filtergraph has been developed by KIS in co-operation with the High Altitude Observatory in Boulder, CO, USA. We acknowledge the use of Gemini to improve the clarity and readability of the text. The authors reviewed, edited, and take full responsibility for the final content of this manuscript.
This research has made use of NASA's Astrophysics Data System.
\end{acknowledgments}

\facilities{ChroTel}

\software{sunpy \citep[][]{2020ApJ...890...68S},
          scikit-image \citep[][]{scikit-image},
          pandas \citep[][]{mckinney-proc-scipy-2010},
          astropy \citep[][]{2013A&A...558A..33A,2018AJ....156..123A}.
          }

\bibliography{biblio}{}
\bibliographystyle{aasjournalv7}







\end{document}